\documentclass{jpsj3}
\usepackage{txfonts}

\title{Bosonic t-J model in a stacked triangular lattice and its phase diagram}

\author{Keisuke Kataoka, Yoshihito Kuno, and Ikuo Ichinose}
\inst{Department of Applied Physics, Nagoya Institute of Technology,
Nagoya, 466-8555 Japan} 

\abst{In this paper, we study phase diagram of a system of
two-component hard-core bosons with nearest-neighbor (NN) pseudo-spin
antiferromagnetic (AF) interactions in a stacked triangular lattice.
Hamiltonian of the system contains three parameters one of which is
the hopping amplitude $t$ between NN sites, and the other two are the NN
pseudo-spin exchange interaction $J$ and the one that measures anisotropy of
pseudo-spin interactions.
Physical states of the system are restricted to the ones without 
the doubly-occupied state at each site.
We investigate the system by means of the Monte-Carlo simulations
and clarify the low-temperature phase diagram.
In particular, we are interested in how the competing orders, i.e., AF order and 
superfluidity, are realized, and also whether supersolid forms as a result
of hole doping into the  state of the $\sqrt{3}\times \sqrt{3}$ pseudo-spin 
pattern with the $120^o$ structure.}

\kword{bosonic t-J model, cold atom, optical lattice, supersolid,
phase separation}

\begin{document}
\maketitle

\section{Introduction}
Study of systems in which competing orders coexist has been one of the
most interesting topics in condensed matter physics.
Among them, 
recent experiments on $^4$He under pressure have led to renewed interest
of supersolid\cite{He}.
Also the search for a lattice supersolid has been motivated by the
realization of optical lattices in ultracold atomic systems.
As dimensionality and interactions between particles are highly controllable
and also there are no effects of impurities, cold atomic system
in an optical lattice is sometimes regarded as a ``final simulator" 
for quantum many-body systems\cite{optical}. 
Numerical studies of hard-core bosons on a triangular lattice find that
stable supersolid states form on doping of holes away from a 
$1/3$-filled solid ($2/3$-filled solid) with the $\sqrt{3}\times \sqrt{3}$ 
pattern\cite{SStri,SS3D}.
This provides example of old idea\cite{AL} that a finite density of defects in the solid
(vacancies or interstitials) may Bose condense and form a superfluid in
the existing crystalline background.
In this paper we shall  pursue the possibility of realization of 
``vacancy condensation"
phenomenon\cite{vacancy} in boson systems in a triangular lattice
in which a frustration exists.

The model that we study in this paper is a system of
two-component hard-core bosons with nearest-neighbor (NN) pseudo-spin
antiferromagnetic (AF) interactions in a stacked triangular lattice.
The particle number at each site is less than unity, and therefore the model is
sometimes called bosonic t-J model\cite{BtJ}, i.e., a bosonic counterpart
of the fermionic t-J model for the strongly correlated electron systems.
The bosonic t-J model on a square lattice was originally studied as an 
effective model of the Bose-Hubbard model with strong on-site repulsions,
and its phase diagram was clarified.
In this paper, we shall study the mode on a stacked triangular lattice.

This paper is organized as follows.
In Sec.2, we shall introduce the bosonic t-J model and explain
its path-integral formalism for the numerical investigation.
In order to incorporate the local constraint faithfully, we introduce a
slave-particle representation.
In Sec.3, we shall show results of the numerical study including 
phase diagrams, various correlation functions, and snapshots.
We find that the system has a rich phase structure.
Physical meanings of the results is discussed by using, e.g., a mean-field
theory like approximation.
Section 4 is devoted for conclusion and discussion.

\section{Model and path-integral formalism}
The mode that we study in this paper is the bosonic t-J model\cite{BtJ} in a
three-dimensional (3D) stacked
triangular lattice whose Hamiltonian is given as follows,
\begin{eqnarray}
H_{\rm tJ}&=&-\sum_{\langle i,j\rangle} t(a^\dagger_{i}a_j
+b^\dagger_{i}b_j+\mbox{h.c.})
+J_z\sum_{\langle i,j\rangle}S^z_{i}S^z_j  \nonumber  \\
&& +J\sum_{\langle i,j\rangle}(S^x_{i}S^x_j+S^y_{i}S^y_j),
\label{HtJ}
\end{eqnarray}
where $a^\dagger_i$ and $b^\dagger_i$ are hard-core 
boson creation operators at site $i$,
pseudo-spin operator $\vec{S}_i={1 \over 2}B^\dagger_i\vec{\sigma}B_i$ with
$B_i=(a_i,b_i)^t$, $\vec{\sigma}$ are the Pauli spin matrices,
and $\langle i,j\rangle$ denotes the nearest-neighbor (NN) sites
in the 3D stacked triangular lattice.
Physical Hilbert space of the system consists of states with total particle
number at each site less than unity 
(the local constraint: $a^\dagger_ia_i+b^\dagger_ib_i\leq 1$).
As we consider the 3D system, there exist finite-temperature ($T$) phase
transitions in addition to ``quantum phase transition", which takes place
as the parameters in $H_{\rm tJ}$ are varied.
The system $H_{\rm tJ}$ might be derived as an effective model of
Bose-Hubbard model that describes a cold atom system in an optical
lattice\cite{Hubbard,BtJ2}.
In this case, $J$ and $J_z$ are related to the intra-species
and inter-species on-site repulsions, and $J<0$.
In this paper we study the bosonic t-J model (\ref{HtJ}) with general interests
and treat $J$ and $J_z$ as free parameters.
We mostly consider the case $J_z, \ J\geq0$, i.e., the frustrated case because
it is very interesting to see how the frustrated pseudo-spin state evolves as 
holes are doped.
Relation between the geometrical frustration and the formation of supersolid 
was discussed previously
for the hard-core bosons in a triangular lattice\cite{SStri,SS3D}.

In addition to its own theoretical interest, another motivation for
studying the model (\ref{HtJ}) with the antiferromagnetic (AF) coupling
is provided by its relation to the fermionic t-J model.
Recently, the fermionic t-J model was studied by mapping it to
a kind of bosonic t-J model by using a Chern-Simons gauge field\cite{CS}.
The resultant bosonic model is closely related to the present model
expressed by using the slave-particle expression given by
Eq.(\ref{slave1})  later on.
In Ref.\cite{CS},  the model was studied by a mean-field type approximation.
Then numerical study of the model (\ref{HtJ}) and to obtain its ``exact
phase diagram" are important for understanding the phase diagram
of the fermionic t-J model.
Effects of the Chern-Simons gauge field will be discussed later on
by comparing the obtained results in this paper and the phase diargam of the
fermionic t-J mode on the triangular lattice\cite{ogata}.
Furthermore very recently, it was shown by the numerical link-cluster
expansion that the Fermi-Hubbard and Bose-Hubbard models on the
square lattice exhibit
very similar behavior in the strong-coupling region\cite{FBHubbard}.
Therefore we expect that the results of the present study gives an important insight
into phase diagram of the fermionic t-J model on the triangular lattice.

In order to incorporate the local constraint faithfully, we use the following
slave-particle representation,
\begin{eqnarray}
&& a_i=\phi^\dagger_i \varphi_{1i}, \;\;\; 
b_i=\phi^\dagger_i \varphi_{2i},  \label{slave}  \label{slave1} \\
&& \Big(\phi^\dagger_i\phi_i+\varphi^\dagger_{1i}\varphi_{1i}+
\varphi^\dagger_{2i}\varphi_{2i}-1\Big)
|\mbox{phys}\rangle =0,
\label{const}
\end{eqnarray}
where $\phi_i$ is a boson operator that annihilates hole at site $i$,
whereas $\varphi_{\sigma i}\ (\sigma=1,2)$ are bosons that represent the pseudo-spin 
degrees of freedom.
$|\mbox{phys}\rangle$ is the physical state of the slave-particle Hilbert space. 
Then the partition function $Z$ at temperature $T$ is given as follows in the
path-integral methods,
\begin{equation}
	Z=\int [D\phi D\varphi_1D\varphi_2]e^{-\beta H_{\rm tJ}},
\label{ZtJ}
\end{equation}
where $\beta=1/(k_BT)$, $H_{\rm tJ}$ is obtained by substituting
the slave-particles representation (\ref{slave}) into Eq.(\ref{HtJ}),
and the path integral is performed with satisfying the slave-particle
constraint (\ref{const}).
The original path-integral representation of the partition function
contains terms like $\bar{a}_i\partial_\tau a_i$, where $\bar{a}_i$
is the complex number corresponding to $a^\dagger_i$ and $\tau$ is the imaginary time.
As we discussed in the previous papers\cite{prev} 
and also showed explicitly by the numerical studies on certain 
models\cite{SS3D,sawa,naka}, 
effect of non-zero Matsubara-frequency modes in the 
{\em 3D system at finite temperature} is mostly
the renormalization of the critical temperature and then the partition function 
in Eq.(\ref{ZtJ}) is a good approximation for studying phase diagram at finite-$T$.
Furthermore, as the system in the 3D stacked lattice can be in a sense regarded as a 
sequence of the 2D system, its low-$T$ phase diagram is closely related 
to that of the 2D system at $T=0$. 
Therefore, it is expected that the low-$T$ phase diagram of 
the 3D system in the stacked
lattice obtained from $Z$ in Eq.(\ref{ZtJ}) is quite similar to that of 
the 2D system at $T=0$ as verified in the previously studied 
cases\cite{SS3D,sawa,naka}.
In other words, {the spatial third direction perpendicular to the 2D 
lattices plays a role similar to the imaginary-time direction}.
More detailed discussion on the model (\ref{HtJ}) at $T=0$
will be given in a future publication.
Furthermore,
it should be remarked here that a closely related discussion on the
appearance of a Lorentz-invariant critical theory was given
for the quantum phase model and the existence of Higgs mode was predicted 
in a two-dimensional superfluid on an optical lattice\cite{Higgs}.

\section{Numerical study}

We employ both the grand-canonical and canonical ensemble for the
practical calculation.
In the grand-canonical ensemble, the chemical potential term like
$\mu \sum_i\phi^\dagger_i\phi_i$ is added to $H_{\rm tJ}$.
On the other hand in the canonical ensemble, the path integral
in Eq.(\ref{ZtJ}) is evaluated by means of the Monte-Carlo simulations
with keeping the average density of holes fixed.
To show results of numerical study, it is convenient to introduce
the following dimensionless parameters, $c_J=\beta J, \ c_{Jz}=\beta J_z$,
$c_t=\beta t$ and $\alpha=J/J_z$.
Therefore large $c_J \ (c_{Jz})$ and/or $c_t$ corresponds to low-$T$ region.

To study the phase diagram, we calculate the internal energy $E$ and 
the specific heat $C$ defined as
\begin{equation}
E=\frac{1}{N}\langle {H}_{tJ} \rangle, \;\;
C=\frac{1}{N}\langle ({H}_{tJ}-E)^2\rangle,
\end{equation}
where $\ N\equiv L^3$ with the linear system size $L$. 
We performed calculation up to $L=30$, and
for the simulations, we employ the standard Monte-Carlo Metropolis algorithm
with local update\cite{Met}.
The typical sweeps for measurement is $(30000 \sim 50000)\times (10$ samples),
and the acceptance ratio is $40\% \sim 50\%$.
Errors are estimated from 10 samples with the jackknife methods.
We also calculate the hole density $\rho=\langle \phi^\dagger_i\phi_i\rangle$,
and correlation functions 
$G_{\rm xy}(i,j)=\langle (S^x_{i}S^x_j+S^y_{i}S^y_j)\rangle$,
$G_{\rm z}(i,j)=\langle S^z_{i}S^z_j\rangle$,
and $G_{\rm S}(i,j)=\langle \vec{S}_{i}\cdot \vec{S}_j\rangle$.
It easily verified $\vec{S}^2_{i}=(a^\dagger_i a_i+b^\dagger_ib_i)^2$,
and then the magnitude of the pseudo-spin is decreased by the 
doping of hole.
On the other hand, 
boson correlation function $G_{\rm B}(i,j)=\langle B^\dagger_{i}B_j\rangle$
is used to see if a superfluid forms.

\begin{figure}[h]
\begin{center}
\includegraphics[width=10cm]{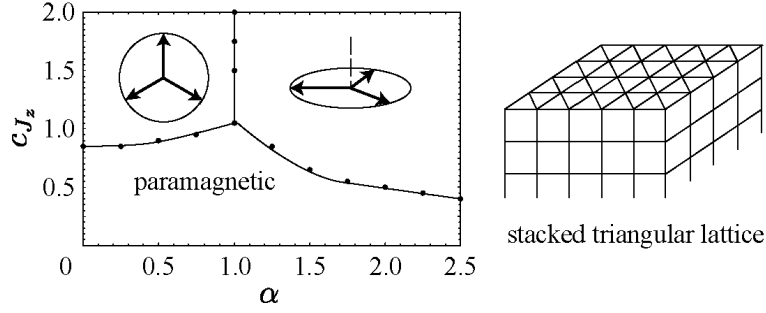} 
\caption{
Phase diagram for anisotropic AF Heisenberg model in a stacked
triangular lattice. 
Vertical axis plots $c_{Jz}$.
At low $T$, there are two phases with a long-range order of 120$^o$
pattern, which are separated with each other by the line $\alpha=1$
($J=\alpha J_z$).
Phase transitions from paramagnetic to ordered states are of second order,
whereas that of $\alpha=1$ is called morphotropic phase boundary\cite{II}.
System size $L=24$ with the periodic boundary condition.
}\vspace{-0.5cm}
\label{PD1}
\end{center}
\end{figure}
\begin{figure}[h]
\begin{center}
\vspace{0.5cm}
\includegraphics[width=5cm]{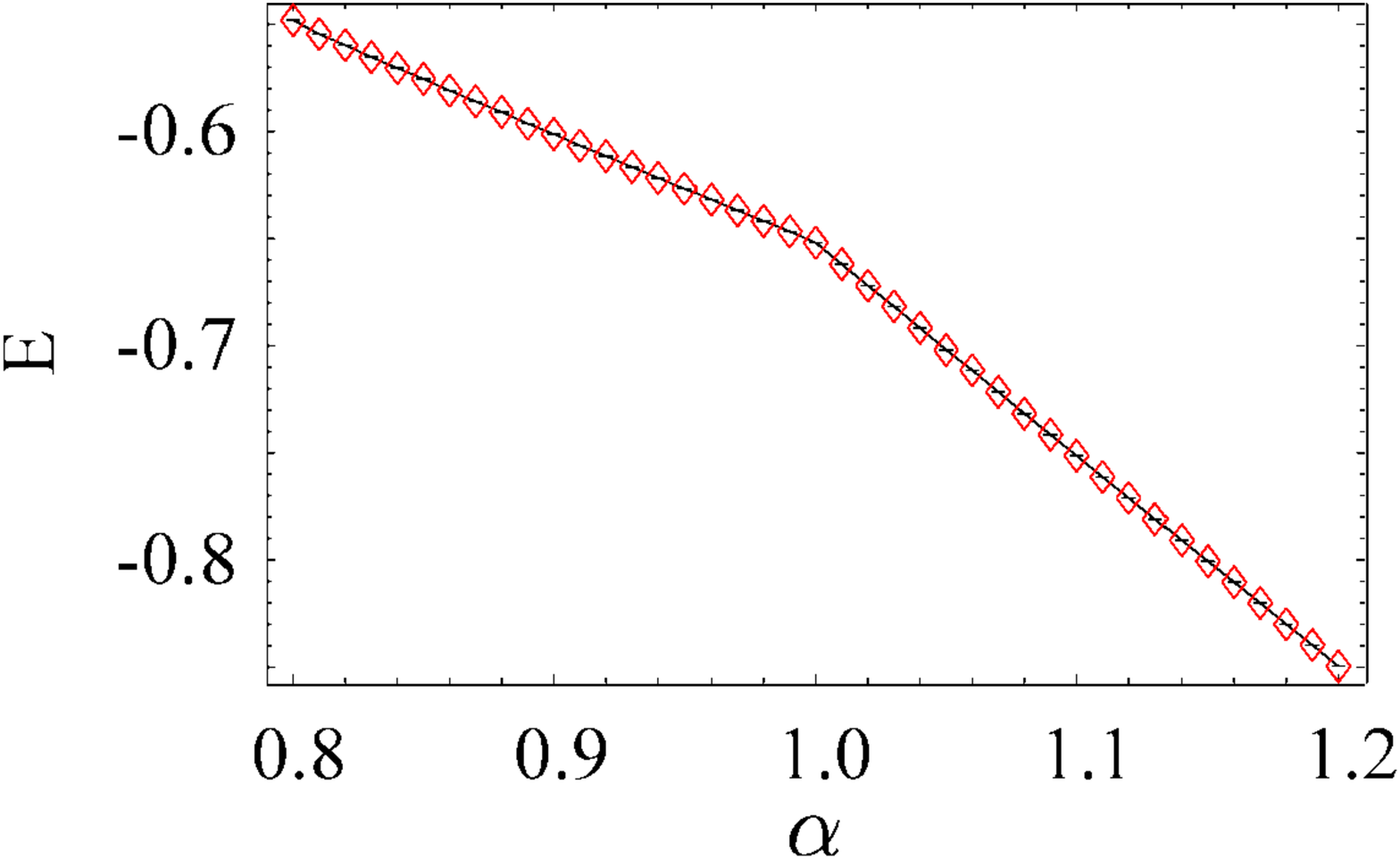}
\hspace{0.5cm}
\includegraphics[width=5cm]{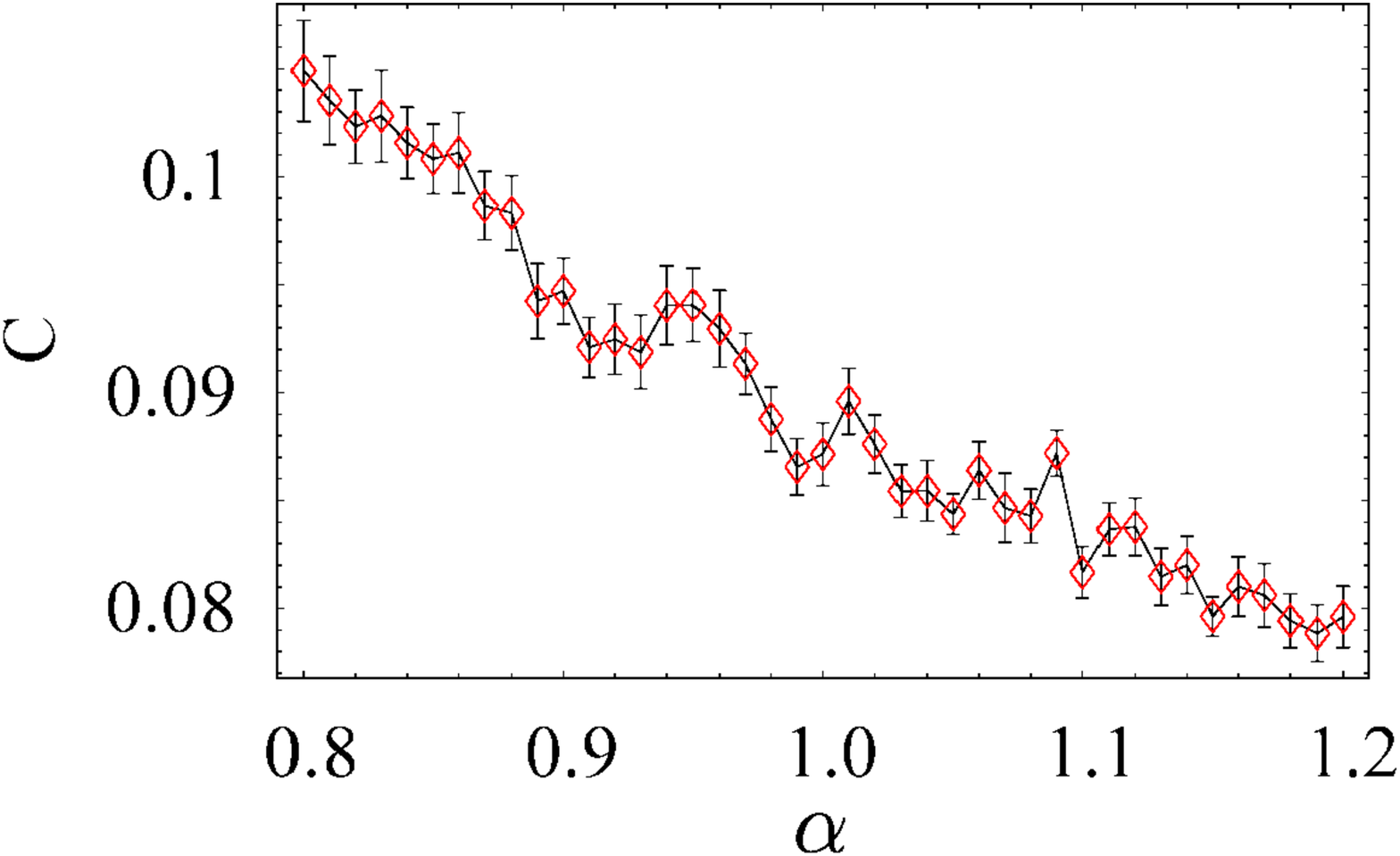}
\vspace{-0.3cm}
\caption{
$E$ and $C$ as a function of $\alpha$ for $c_{Jz}=1.5$.
At $\alpha=1$, $C$ does not have a peak but exhibits some 
peculiar behavior though $E$ indicates the existence of the phase transition.
This is a typical behavior of the morphotropic phase boundary.
}\vspace{-0.5cm}
\label{alpha}
\end{center}
\end{figure}
\subsection{Anisotropic AF Heisenberg model on the stacked triangular lattice}

At zero hole density, the system (\ref{HtJ}) reduces to the anisotropic 
AF Heisenberg model.
The same model in the 2D triangular lattice has been studied rather
intensively\cite{2Dtri}.
To identify the phase boundary of the present 3D model, 
$E$ and $C$ were calculated by the Monte-Carlo simulations. 
The spin correlation functions were also calculated in order to identify
each phase.
We show the obtained phase diagram in Fig.\ref{PD1}.
As $T$ is lowered, second-order phase transition from the paramagnetic phase to 
the spin-ordered states takes place.
Specific heat $C$ exhibits a sharp peak at the transition points indicating 
the existence of the second-order phase transition.
Location of the phase transition points is determined by the peaks of $C$.
In the low-$T$ region,  there are two phases separated with each other by 
the line $\alpha=1$.
In the study of ferroelectric materials, the corresponding line
is called the morphotropic phase boundary, and it plays an important role\cite{II}.
We show the internal energy $E$ and the specific heat $C$
as a function of the anisotropy $\alpha$ in Fig.\ref{alpha},
which exhibit a ``typical" behavior of the morphotropic phase boundary\cite{BtJ2}.
It should be remarked that the system has the $SU(2)$ 
pseudo-spin symmetry at $\alpha=1$, whereas $U(1)\times Z_2$ otherwise.
This fact is the origin of the peculiar behavior of $E$ and $C$ at $\alpha=1$.
The phase structure at low $T$, i.e., $c_J, \ c_{Jz} \rightarrow$ large, 
is essentially the same with that of the
2D system at $T=0$, as it is expected from the above general consideration.
We have also studied the correlations of spins in the vertical direction of
the stacked triangular lattice, and found the strong AF long-range correlations
in the two phased with the 120$^o$ spin order and short-range one in 
the paramagnetic phase in Fig.\ref{PD1}.

\begin{figure}[h]
\begin{center}
\includegraphics[width=6cm]{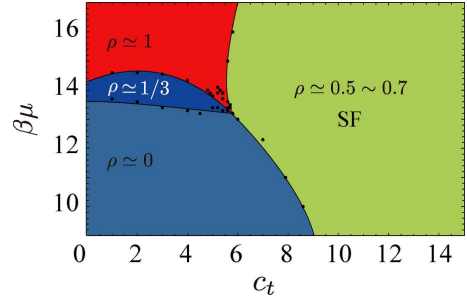}
\vspace{-0.3cm}
\caption{
Phase diagram for bosonic t-J model in a stacked
triangular lattice at low $T$, $c_J=6.0$ (grand-canonical ensemble).
$\rho$ is hole density.
Dots denote location of the phase transition observed by the numerical
study.
All phase transitions are of first order. SF stands for superfluid
and this phase also has FM long-range order.
The phase $\rho\simeq 0$ is essentially the pure spin system
with the $\sqrt{3}\times \sqrt{3}$ symmetry and
the phase with $\rho\simeq 1$ is the empty state.
The phase with $\rho\simeq 1/3$ is explained in Fig.\ref{1/3_grand}.
These three phases are all insulating.
}\vspace{-0.5cm}
\label{PD2}
\end{center}
\end{figure}
\begin{figure}[h]
\begin{center}
\vspace{0.5cm}
\includegraphics[width=5cm]{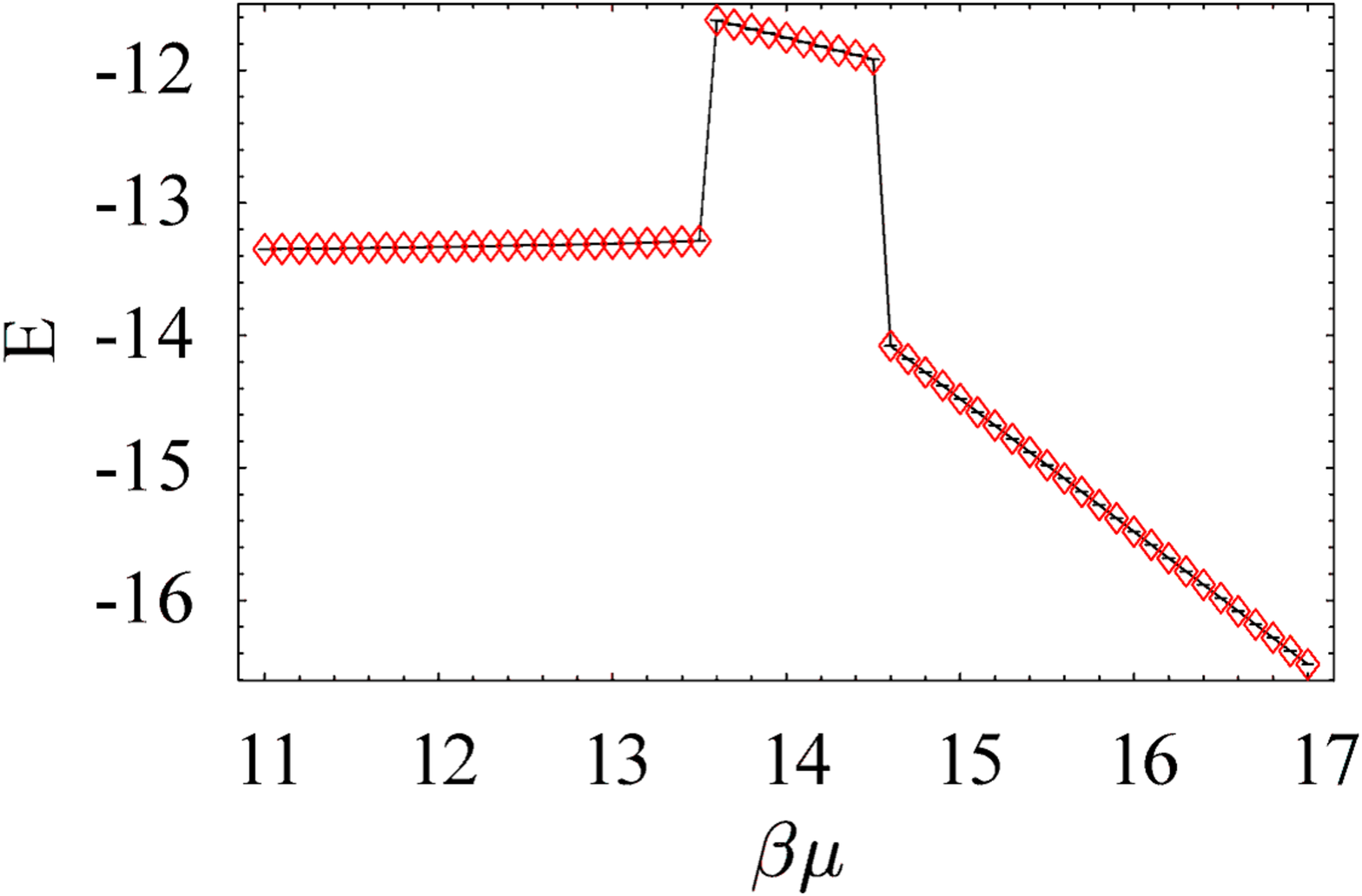}
\includegraphics[width=5cm]{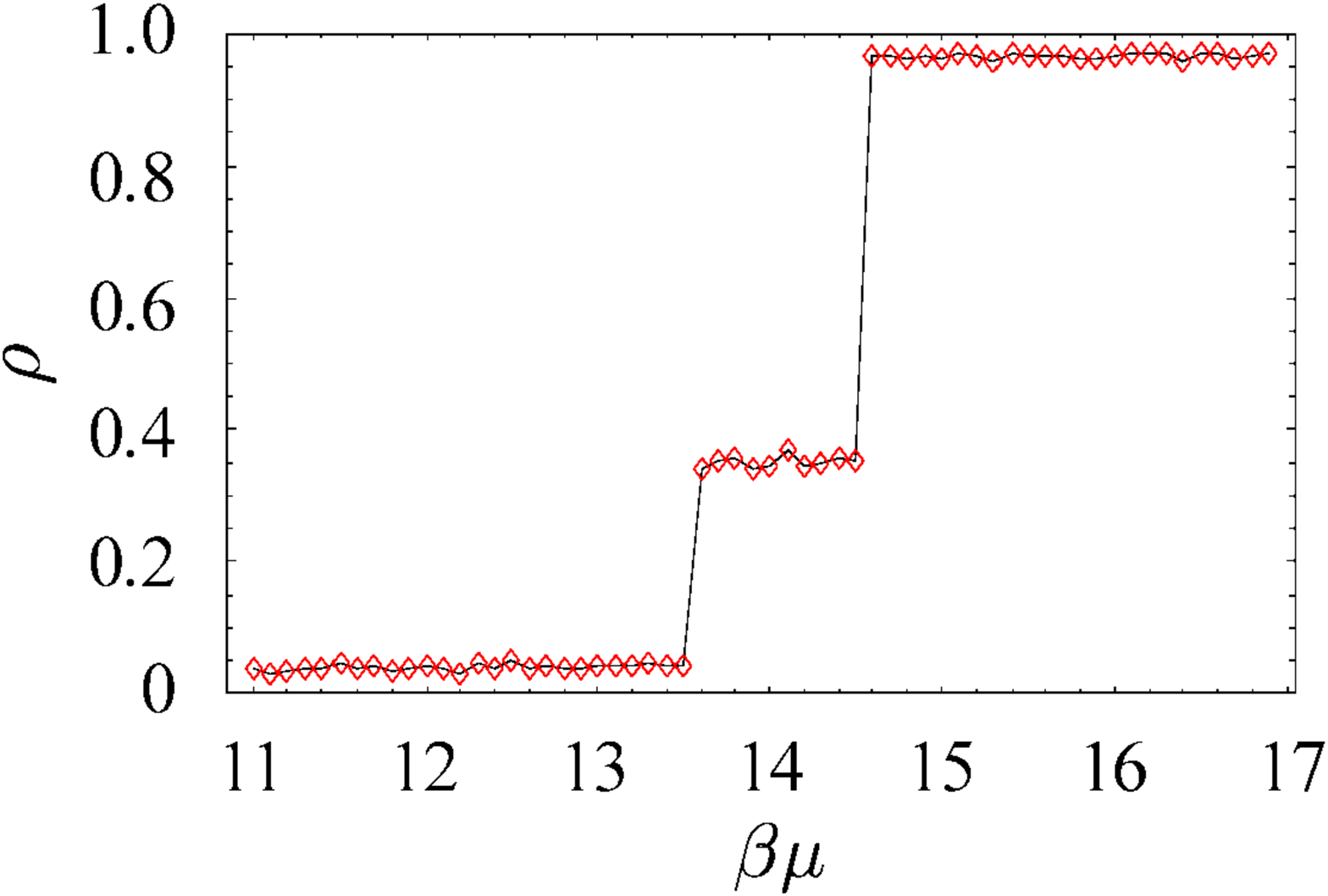}
\vspace{-0.3cm}
\caption{
$E$ and $\rho$ as a function of $\mu$ with $c_J=6.0$ and $c_t=2.0$.
There are two first-oder phase transitions.
$L=30$.
}\vspace{-0.5cm}
\label{Erho}
\end{center}
\end{figure}
\begin{figure}[h]
\begin{center}
\includegraphics[width=5cm]{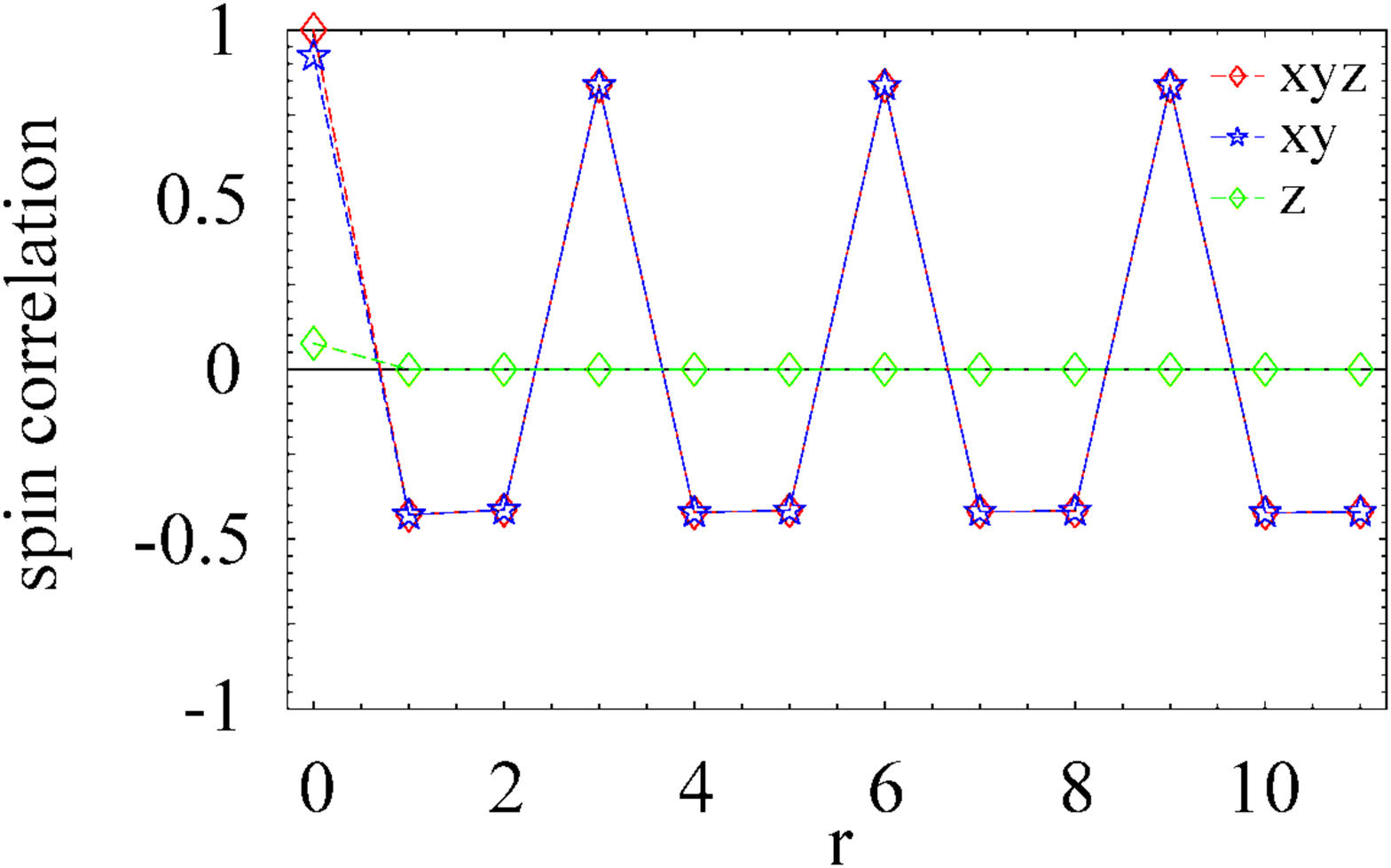}
\includegraphics[width=5cm]{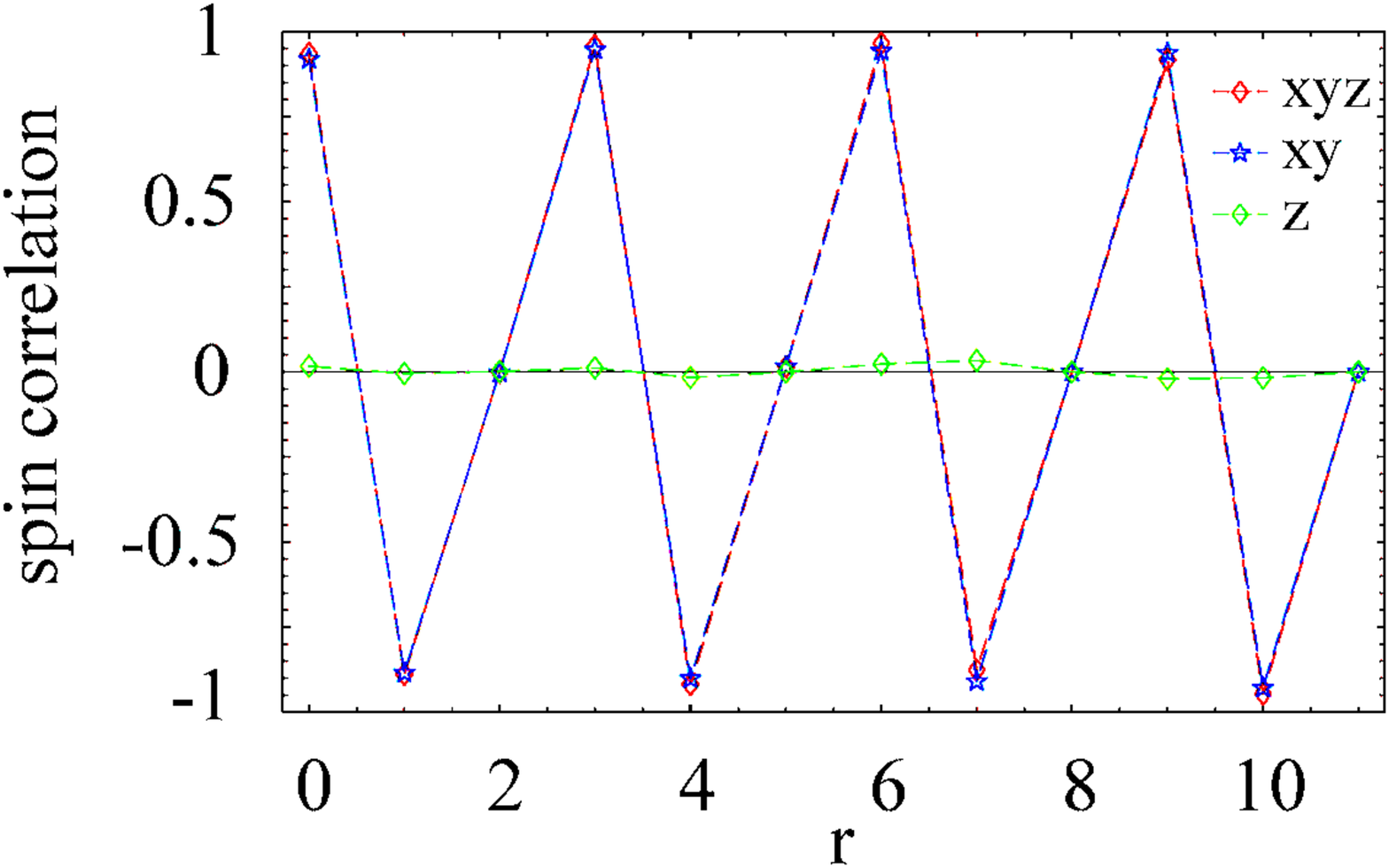}
\vspace{-0.3cm}
\caption{
Spin correlation functions $G_{\rm S}(i,j),  G_{\rm xy}(i,j)$
and $G_{\rm z}(i,j)$ for $c_t=2.0$ and $\beta\mu=12.0$ (left),
$\beta\mu=14.0$ (right).
}\vspace{-0.5cm}
\label{spinC}
\end{center}
\end{figure}

\subsection{t-J$_{xy}$ model}

In this subsection, we focus on the $xy$ AF Heisenberg model
by setting $J>0, \ J_z=0$ in Eq.(\ref{HtJ}) and study effect of the hole doping.
The case of $J=0,\ J_z>0$ will be investigated in the subsequent subsection.
For the ferromagnetic case $J<0$ and $|J|>|J_z|$, it is readily expected that the
system has a similar phase diagram to that on the cubic lattice\cite{BtJ2}
as there are no frustrations in the system.
One of the motivations to study the doped AF magnets on the triangular lattice
comes from the possible spin liquid and superconductivity suggested by
Anderson\cite{RVB}, and therefore we focus on the AF cases.

We first investigate the system in the grand-canonical ensemble.
Obtained phase diagram in the $c_t-\beta\mu$ plane for $c_J=6.0$
is shown in Fig.\ref{PD2}.
For the region $c_t <5.5$, there exist three phases
and they are separated by sharp first-oder phase transitions.
See calculations of $E$ and the averaged hole density
$\rho=\langle \phi^\dagger_i \phi_i\rangle$ in Fig.\ref{Erho}, the both of which
exhibit sharp discontinuities at $\beta\mu\simeq 13.5$ and $14.7$.
From this observation, we judge the existence of the first-order
phase transitions there.
The location of the phase transition points is determined by $E$.

\begin{figure}[t]
\begin{center}
\includegraphics[width=4.5cm]{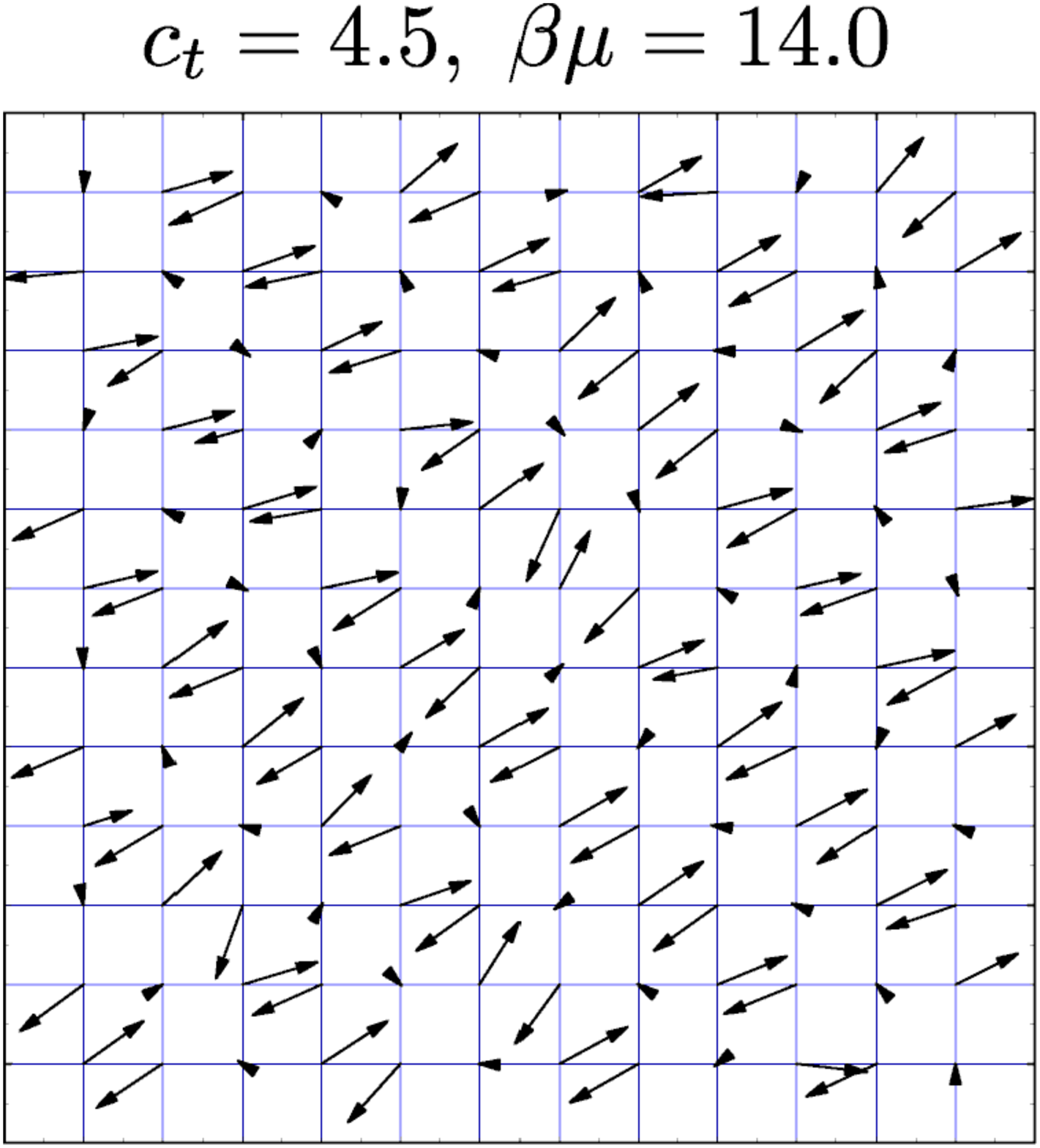}
\hspace{0.2cm}
\includegraphics[width=3cm]{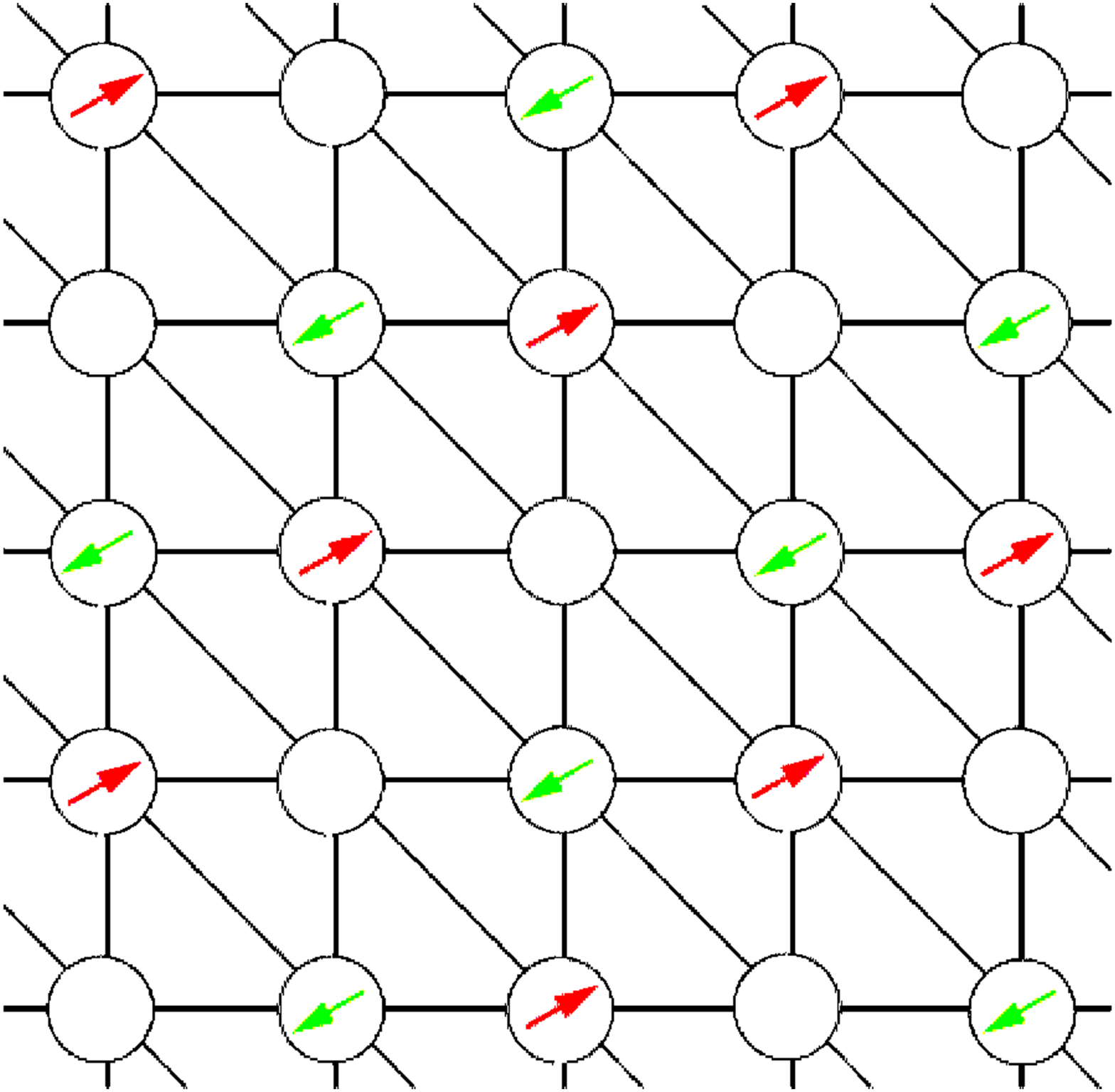}
\caption{
(Left) Snapshot for $c_t=4.5$ and $\beta\mu=14.0$.
Holes are localized and a kind of AF configuration of pseudo-spin
is realized.
We show pseudo-spins in the $S^x-S^y$ plane for $S^z$ component is
negligibly small.
Therefore the length of arrows indicates magnitude of pseudo-spins.
(Right) Caricature of  typical configuration obtained by MC simulation.
}\vspace{-0.5cm}
\label{1/3_grand}
\end{center}
\end{figure}

It is obvious that the phase for $\beta\mu<13.6$ is nothing but
the pure spin system of very low hole density that has the 
long-range order of the three-sublattice $\sqrt{3}\times \sqrt{3}$ pattern.
See the correlation function for $\beta\mu=12.0$ in Fig.\ref{spinC}.
On the other hand in the intermediate region $13.6 <\beta \mu <14.4$,
stable state with $\rho={1 \over 3}$ is realized.
Correlation functions in Fig.\ref{spinC} indicate that the state is nothing but the one 
shown by the snapshot in Fig.\ref{1/3_grand}.
We verified this conclusion by calculating the hole density. 
For $\mu>14.4$ the density of hole is almost unity and the empty state 
forms there.

\begin{figure}[t]
\begin{center}
\includegraphics[width=5cm]{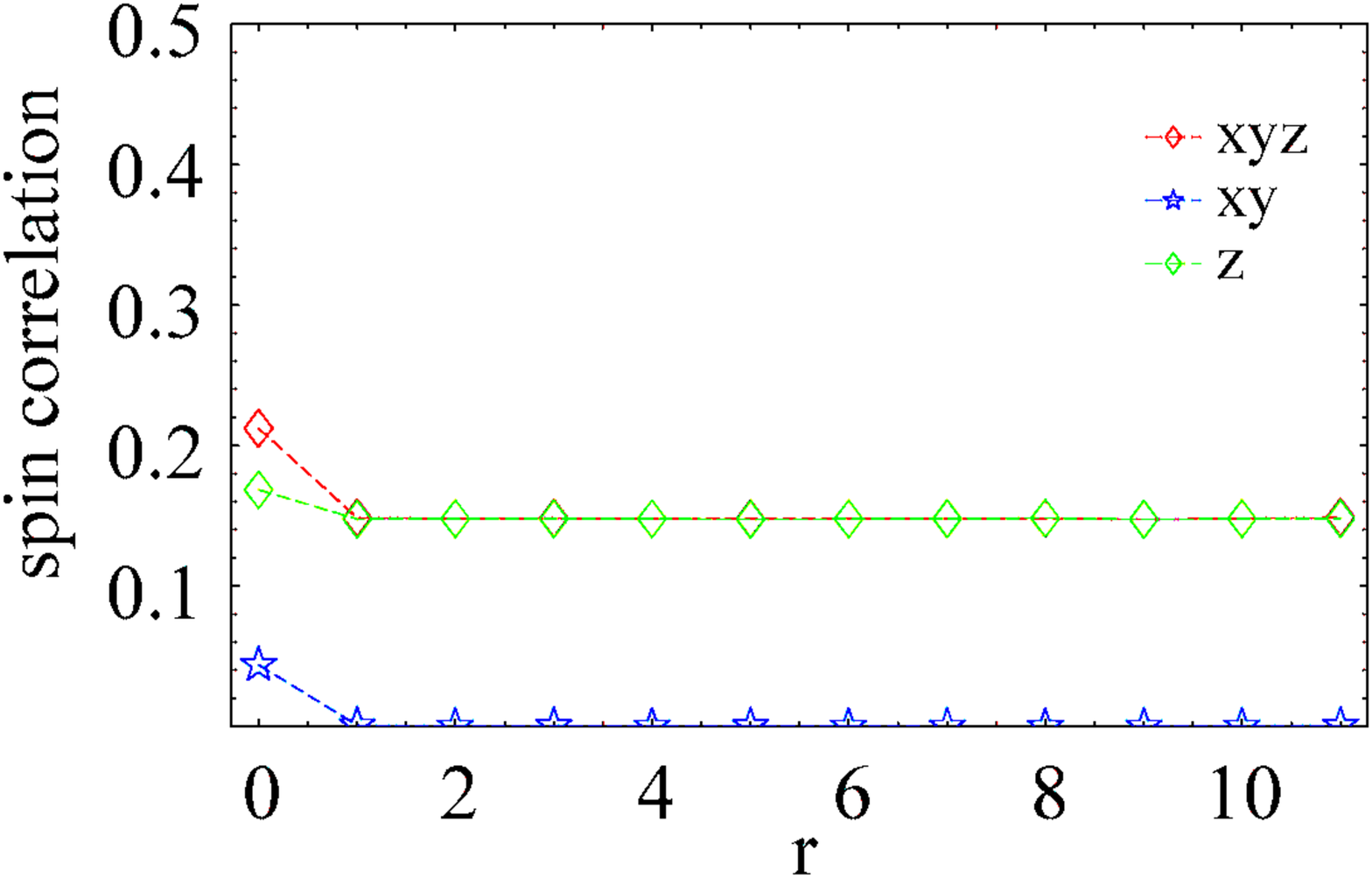}
\hspace{0.5cm}
\includegraphics[width=5.4cm]{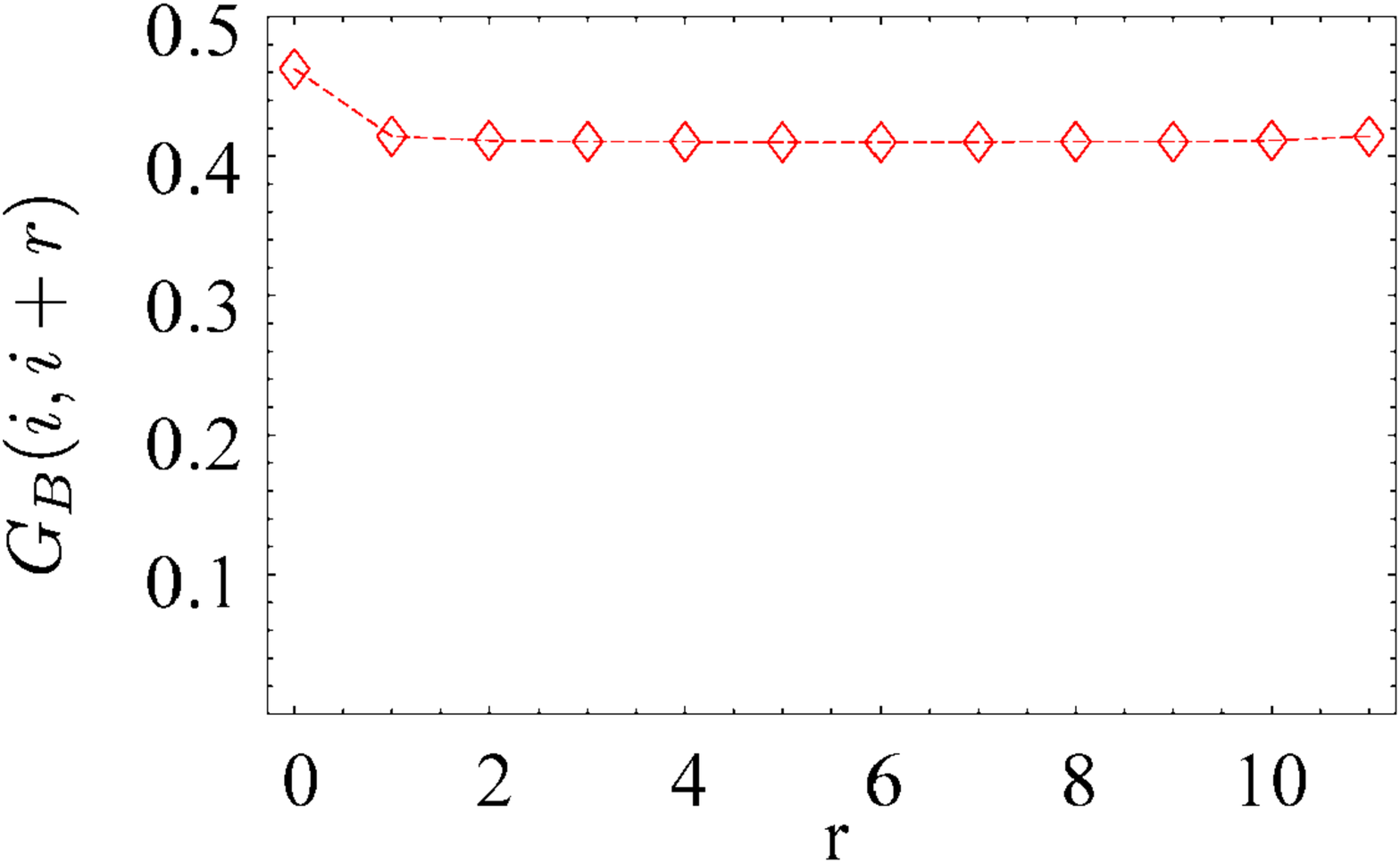}
\vspace{-0.3cm}
\caption{
Spin (left) and boson (right) correlation functions for $c_t=12.0$ and 
$\beta\mu=11.0$ 
}\vspace{-0.5cm}
\label{correlation2}
\end{center}
\end{figure}

As $c_t$ is increased, all the above three phases make a phase transition
to a new phase via first-order phase transitions.
$E$ exhibits discontinuities at the phase boundary of the phase 
$\rho\simeq 0.5\sim 0.7$ in Fig.\ref{PD2}, indicating first-order
phase transitions.
Hole density of this phase is $\rho \simeq 0.5-0.7$, and the calculated spin correlation
indicates the existence of the ferromagnetic (FM) long-range orders 
$\langle S^z_i\rangle \neq 0$ and 
also non-vanishing superfluidity $\langle B_i\rangle \neq 0$. 
See Fig.\ref{correlation2}.
This result means that the system is composed of, say, $a$ bosons and
the Bose condensation of $a$ boson takes place there.
If we impose the condition that the total number of $a$ boson and that of
$b$ boson are equal, the phase separation to $a$-rich region and $b$-rich
region is expected to occur.
This problem is under study and result will be published in a near future.

In the grand-canonical ensemble, the most stable state in the system appears for
each value of the chemical potential.
Near the first-order phase transition, it is rather difficult to control the
particle density by varying the value of the chemical potential.
Then it is quite interesting to study the system in the canonical ensemble by
keeping the hole density constant.
In the present system, we focus on how the state of hole density, say, 40\% evolves 
as the hopping parameter $t$ is increased.

\begin{figure}[h]
\begin{center}
\includegraphics[width=6cm]{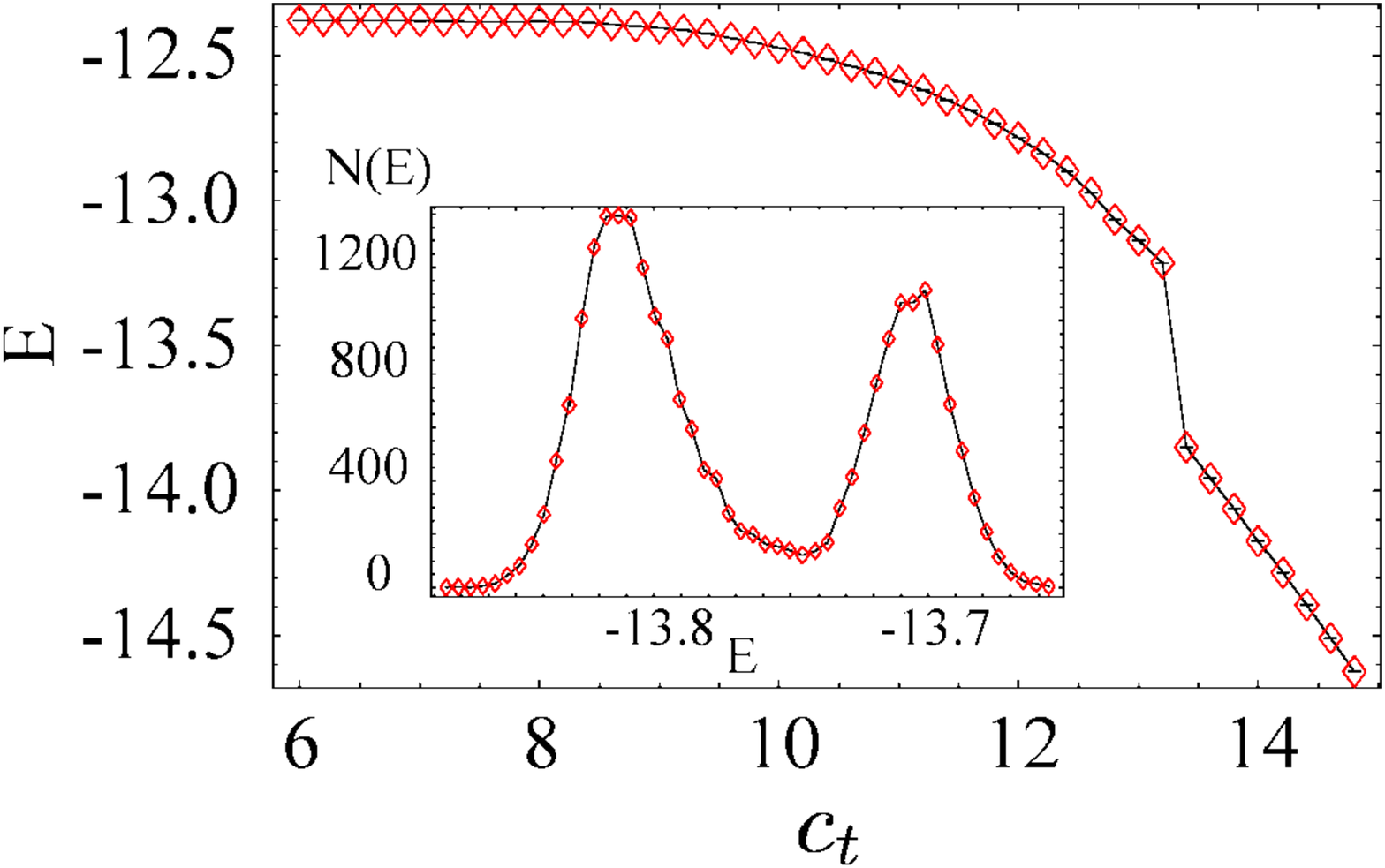}
\vspace{-0.3cm}
\caption{
Internal energy $E$ as a function of $c_t$ for $\rho=0.4$ and $c_J=10.0$.
At $c_t\simeq 13$, there exists a first-order phase transition.
(Inset) $N(E)$ for $c_t=13.2$ exhibits the double-peak shape
indicating the first-order phase transition.
}\vspace{-0.5cm}
\label{E_0.4}
\end{center}
\end{figure}
\begin{figure}[b]
\begin{center}
\includegraphics[width=4cm]{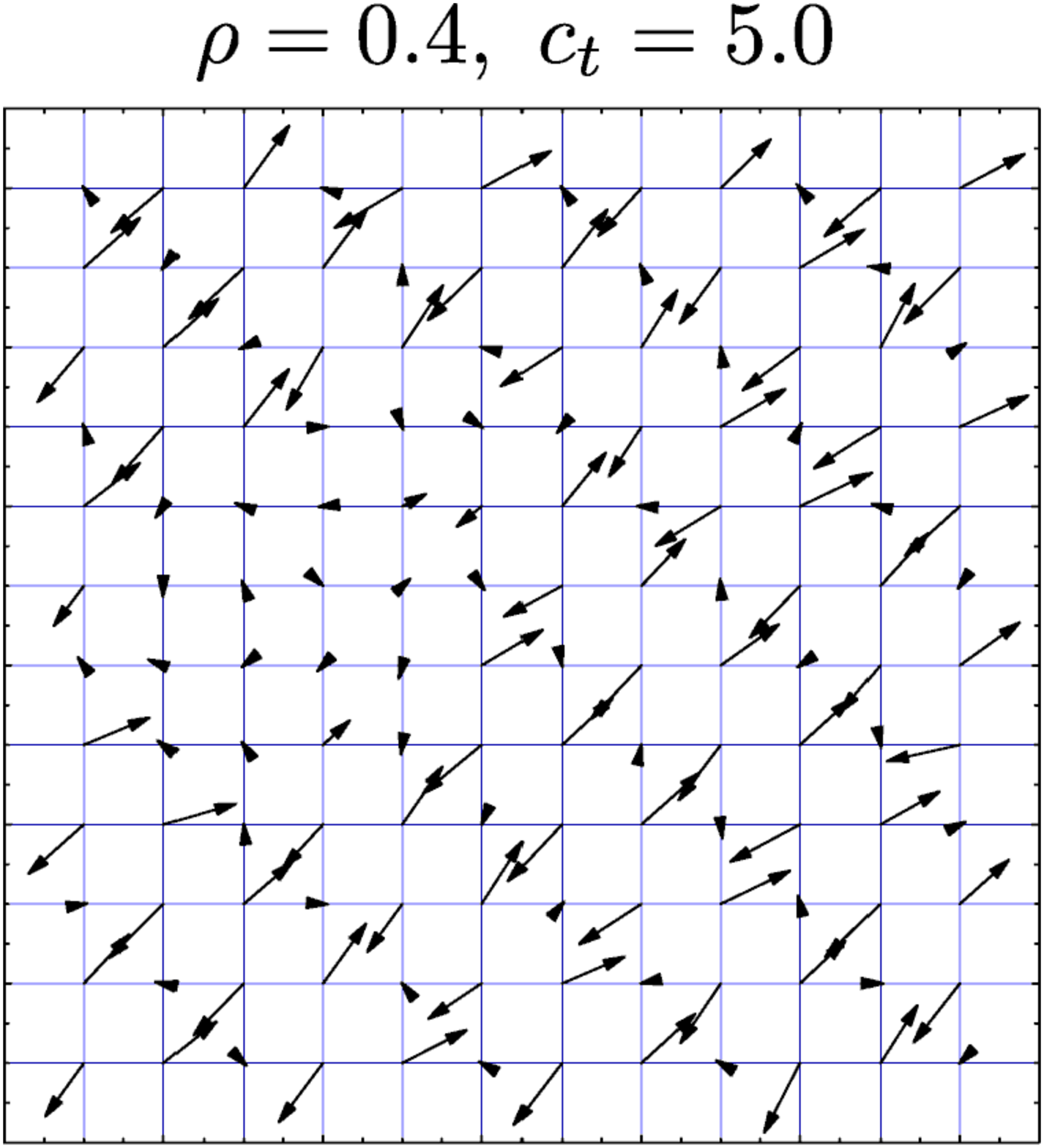}
\hspace{0.5cm}
\includegraphics[width=4cm]{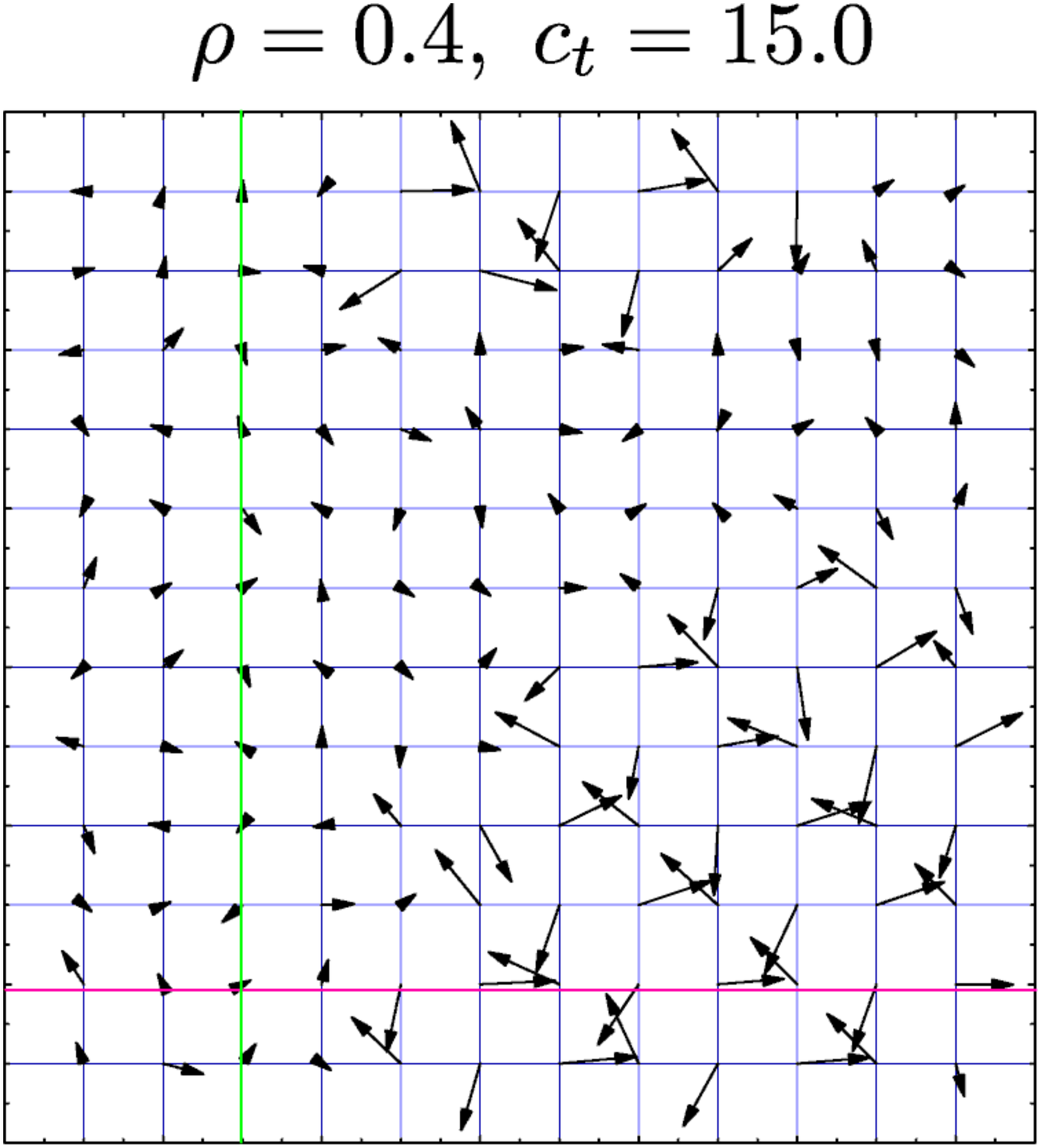}
\hspace{0.5cm}
\includegraphics[width=5.5cm]{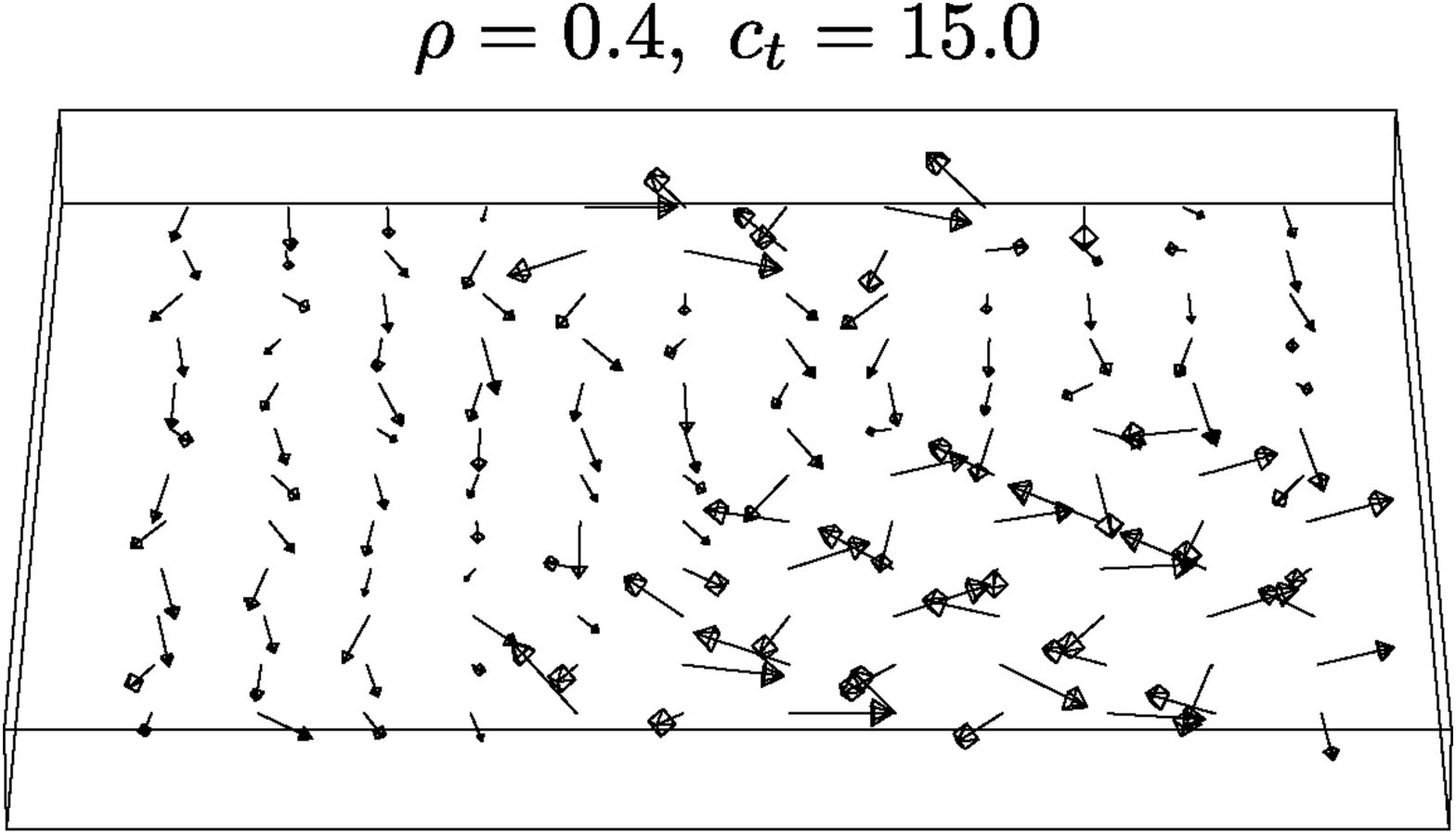}
\vspace{-0.3cm}
\caption{
Snapshots for $\rho=0.4,\ c_J=10.0$ and $c_t=5.0$ (left), 
$c_t=15.0$ (center, right (oblique angle)). 
Left and center snapshots show $S^x-S^y$ component of pseudo-spin.
Right one is snapshot from an oblique angle, and
length of arrows indicates magnitude of pseudo-spins.
}\vspace{-0.5cm}
\label{snapshots}
\end{center}
\end{figure}

\begin{figure}[h]
\begin{center}
\includegraphics[width=5cm]{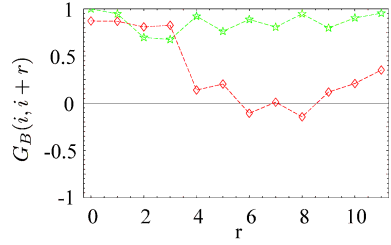}
\vspace{-0.3cm}
\caption{
Boson correlation functions for $\rho=0.4$ and $c_t=15.0$ along the two lines
shown in Fig.\ref{snapshots}. 
}\vspace{-0.5cm}
\label{correlationB}
\end{center}
\end{figure}

In Fig.\ref{E_0.4}, we show the internal energy $E$ for $c_J=10.0$ as a function of $c_t$.
We also measures the number of states $N(E)$, which is defined as 
$Z=\int dE N(E)\exp(-\beta E)$.
The result indicates that there exist a first-order phase transition at $c_t=13.2$.
In order to understand the physical meaning of the phase transition, it is
quite useful to see snapshots of the two phases separated by the phase transition.
See Fig.\ref{snapshots}.
From the snapshot for $c_t=5$, it can be seen that the phase of 
$\rho={1 \over 3}$ survives and there is a void of very low particle density
as a result of an excess of holes compared to $\rho={1 \over 3}$.
On the other hand, the snapshot for $c_t=15$ shows that the phase separation
takes place, i.e., the region of pure-spin phase with the $\sqrt{3} \times \sqrt{3}$
pattern and the region of the superfluid coexist, but they are immiscible.
In the superfluid region, the boson has a nonvanishing expectation value 
$\langle B_i \rangle \neq 0$.
The observation obtained through the snapshots is verified by the correlation functions
$G_{\rm B}(i,j)$ shown in Fig.\ref{correlationB}.

\begin{figure}[h]
\begin{center}
\includegraphics[width=6cm]{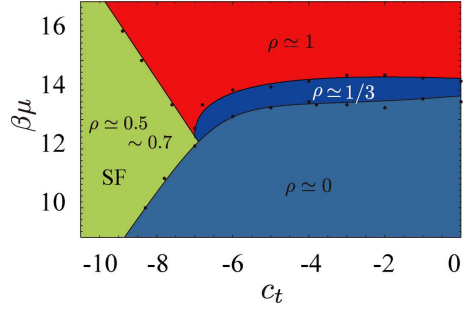}
\vspace{-0.3cm}
\caption{
Phase diagram for $c_t<0$. 
There are four phases as in the case of $c_t>0$ and they are separated
by first-order phase transition lines.
Physical meaning of each phase is explained in the text.
}\vspace{-0.5cm}
\label{PD3}
\end{center}
\end{figure}

\begin{figure}[b]
\begin{center}
\includegraphics[width=5cm]{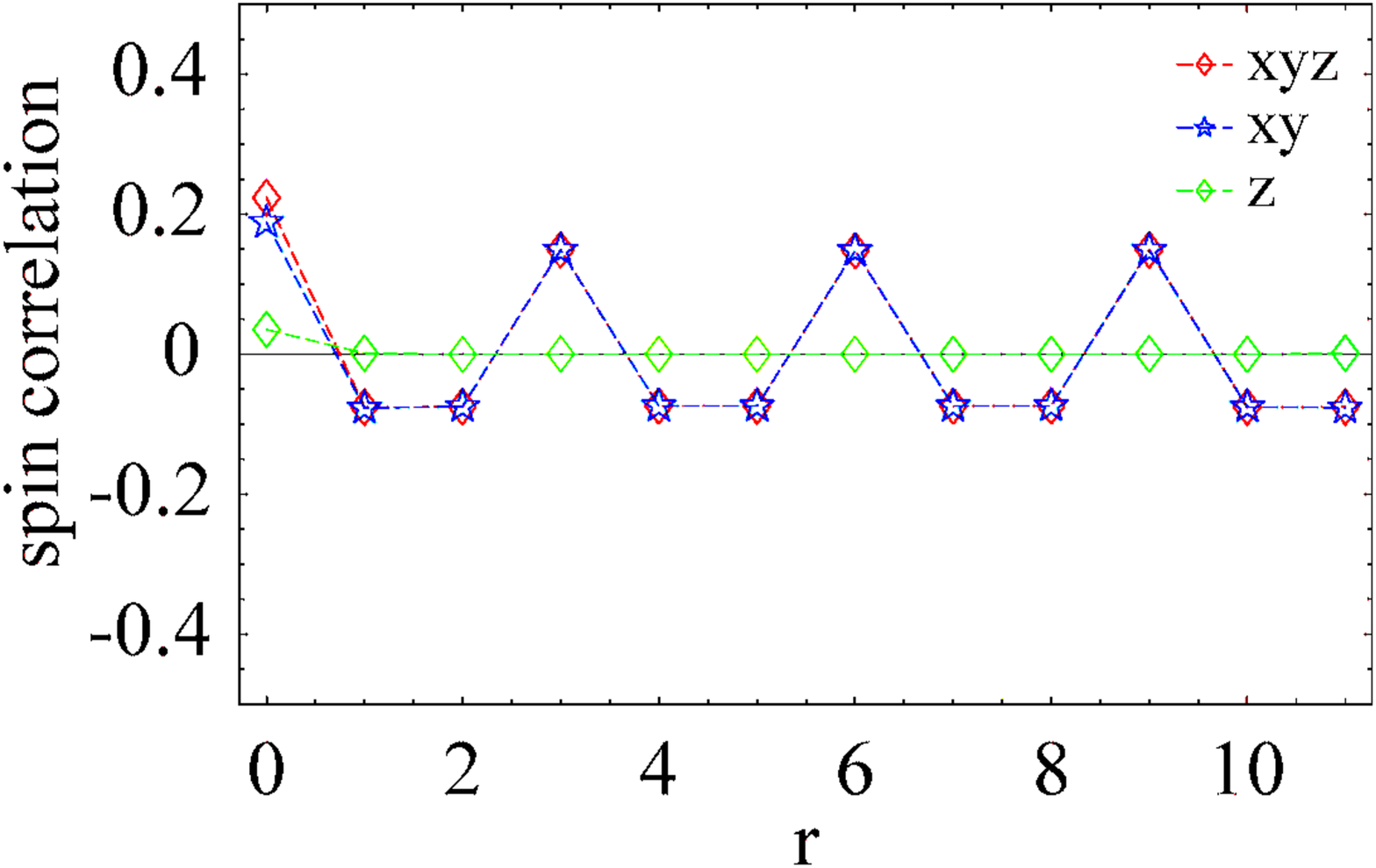}
\hspace{0.5cm}
\includegraphics[width=5cm]{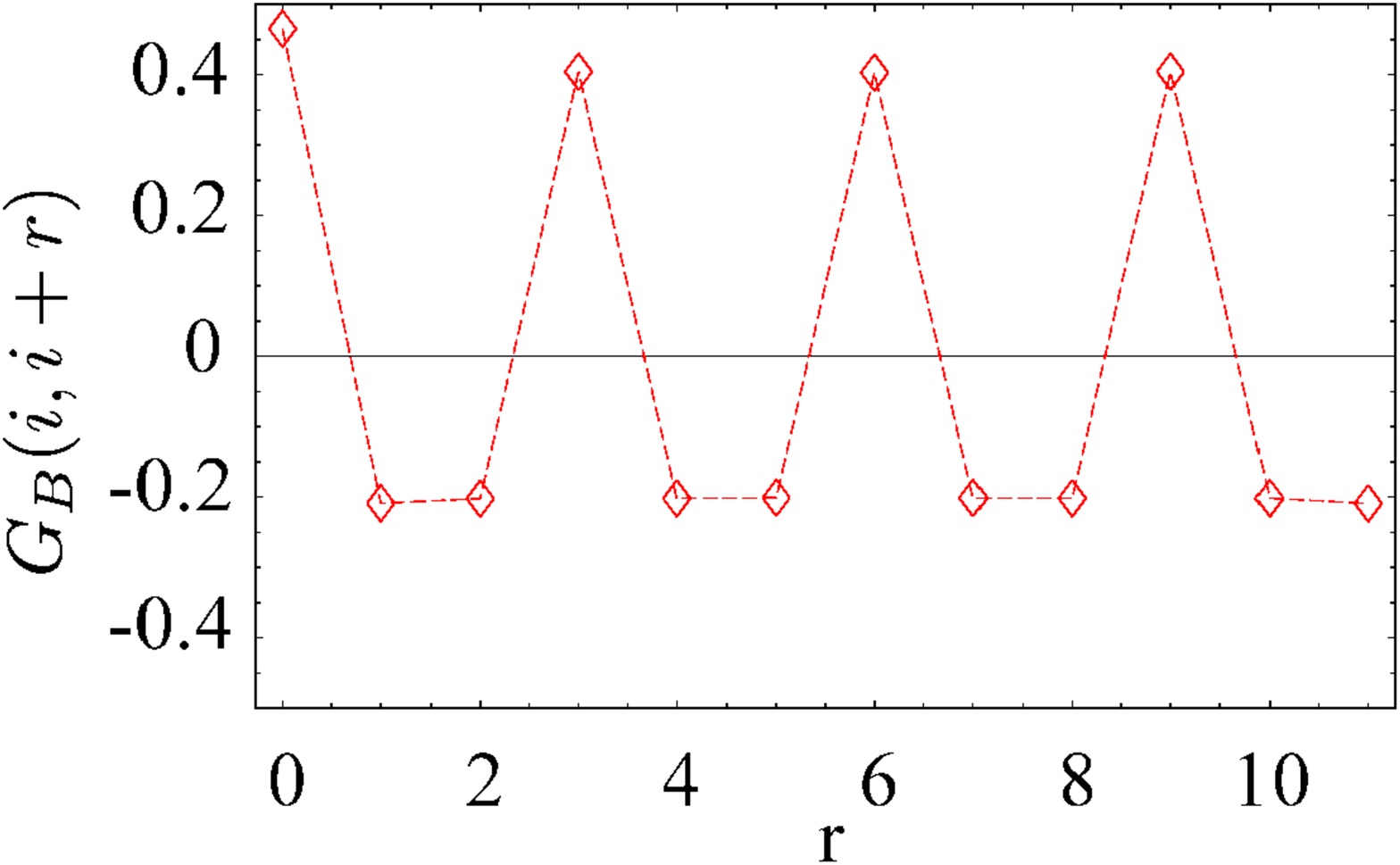}
\vspace{-0.3cm}
\caption{
Spin (left) and boson (right) correlation functions for $c_t=-19.0$ and 
$\beta\mu=11.0$.
The spin correlation function $G_{\rm xy}(i,j)$ exhibits the
$\sqrt{3}\times \sqrt{3}$ pattern.
The boson correlation $G_{\rm B}(i,j)$ also shows a similar behavior.
}
\label{correlation3}
\end{center}
\end{figure}

The above result shows that the phase separation takes place and the
supersolid does not form.
The reason why the first-order phase transition takes place and
the phase-separated state forms in the present model is 
understood as follows.
Bose condensation $\langle B_i\rangle \neq 0$ naturally induces
a spin order.
In the mean-field approximation, the wave function $\Psi_{\rm BC}$ 
of Bose-condensed state with a {\em coplaner spin order in the 
$S^x$-$S^y$ plane} is given as 
$$
\Psi_{\rm BC}\propto
\prod_i [e^{i\eta_i}a^\dagger_i+e^{i\theta_i}b^\dagger_i+c]|0\rangle,
$$
where $c$ is a positive number.
Then $\langle S^x_i \rangle/\langle S^y_i \rangle=\cot(\eta_i-\theta_i)$.
On the other hand, 
$\langle a_i \rangle=ce^{i\eta_i}$ and $\langle b_i \rangle=ce^{i\theta_i}$.
Therefore if the supersolid with the spin $120^o$ long-range order forms, 
the phase of the  superfluid cannot be uniform. 
As a result, the lowest hopping-energy state of the Bose condensate cannot 
be realized.
More precisely, 
the expectation value of the Hamiltonian in the state $\Psi_{\rm BC}$
is evaluated as 
$$
\langle H_{\rm tJ} \rangle_{\rm BC}\sim
-tc^2(\cos \Delta \eta+\cos \Delta \theta)+J\cos (\Delta \eta-\Delta\theta)
$$
where $\Delta \eta$ etc are the phase differences between Bose
condenses on adjacent sites.
In order to generate the Bose condensation, the parameter $t$ has to
exceed some critical value.
The hopping term with the coefficient $t$ prefers
$\Delta \eta,\ \Delta \theta\sim 0$, i.e.,
the Bose condensation tends to accompany a ferromagnetic order.
Then as $c_t$ is increased, a first-oder phase transition takes place and
the system tends to phase separate into the superfluid region of
intermediate particle density and the pure-spin region with $120^o$ spin 
order and $\rho\simeq 0$.

From the above discussion, it is interesting to study the case $c_t=\beta t<0$, 
which is sometimes called frustrated NN hopping.
From the above consideration, one can expect that a state with both a non-colinear 
spin order and the superfluidity with a nonvanishing momentum
(i.e., $\pi>|\Delta \eta|, \ |\Delta \theta|>{\pi \over 2}$) 
forms at sufficiently low $T$. 
Here it should be mentioned that the case $c_t=\beta t<0$ can be realized
in a rotating Bose gas system as rotation of optical lattice generates
an effective magnetic field for bosons\cite{rotation}.
In the case in which magnetic flux penetrating each triangular plaquette is
exactly $\pi$, the frustrated NN hopping is realized.
Furthermore in the fermionic t-J model, the model with $t<0$ describes
the electron doped materials.
As we mentioned before, the fermionic t-J model is related to
its bosonic counterpart through a Chern-Simons gauge theory.
Possible relation of the bosonic t-J model to the fermionic
t-J model will be discussed later on.

We numerically studied the t-J model with a negative $t$ as in the
previous case with $t>0$.
We first show the obtained phase diagram for $c_t<0$ in Fig.\ref{PD3}.
There are four phases and they are separated by first-order phase transitions
as in the previous cases.
The phase transition lines are determined by the measurement of $E$.
The phases with $\rho\simeq 0, {1 \over 3}, 1$ are essentially the same with
the ones shown in Fig.\ref{PD2}.
The new phase that appears for $c_t<-7$ is the expected to have
both the co-planer long-range spin order and the superfluidity.
To see it, we show the spin and boson correlation functions
in Fig.\ref{correlation3}.
It is obvious that the pseudo-spin has the long-range order with 
$\sqrt{3}\times\sqrt{3}$ pattern and also the Bose condensation
with a nonvanishing momentum forms there.
Appearance of this phase comes from the fact that the Bose
condensates $\langle a_i\rangle$ and  $\langle b_i\rangle$ have
different phases depending on the $A, \ B$ and $C$ sublattices,
and these position-dependent condensations are enhanced by both
the 120$^o$ structure of the spin oder and the negative hopping $t<0$.
This the reason why this phase is different from that for large positive
$t$ in Fig.\ref{PD2}.

The correlation function $G_{\rm z}(i,j)$ in 
Fig.\ref{correlation3} exhibits 
vanishing value and therefore the density of atoms are uniform even for $c_t<-7$.
This means that supersolid does not form in the present system.
In the following subsection, we shall study the t-J$_z$ model and 
show existence of the supersolid phase as a result of the competition of
$t$ and $J_z$ terms.
As the t-J$_z$ model with $J_z<0$ is directly derived from the Bose-Hubbard model
with repulsions, the obtained phase diagram is expected to be verified by experiments
of the two-component cold atom systems. 

\subsection{t-J$_z$ model}

In the previous subsection, we studied the t-J$_{xy}$ model and
clarified its phase diagram.
There we found that there the supersolid does not form though the 
phase with the both the spin and boson long-range orders exists in
some parameter region.
In this subsection, we shall continue the numerical study on the bosonic t-J model 
and consider the case with $J_z>0, \ J=0$.
The case $J_z>0$ is realized in the two-component cold atom system
in which the intra-species repulsion is larger than the inter-species 
one\cite{BtJ2}.
As $S^z_i={1 \over 2}(a^\dagger_i a_i-b^\dagger_i b_i)$, the existence of 
the long-range orders $\langle S^z_i \rangle \neq 0, \langle a_i \rangle \neq 0$ 
$(\langle b_i \rangle \neq 0)$ induces a genuine supersolid.
\begin{figure}[h]
\begin{center}
\includegraphics[width=7cm]{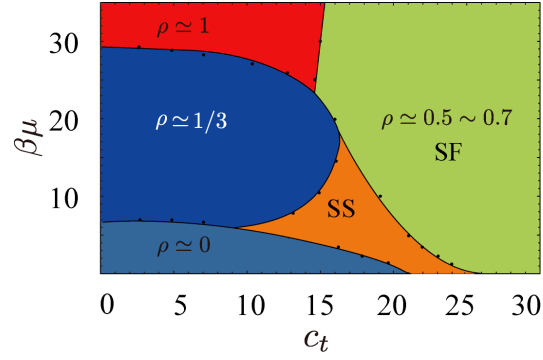}
\vspace{-0.3cm}
\caption{
Phase diagram of t-J$_z$ model in the grand-canonical ensemble.
There are five phases and they are separated
by first-order phase transition lines.
SS stands for supersolid.
$c_{Jz}=10.0$.
}\vspace{-0.5cm}
\label{PD4}
\end{center}
\end{figure}
\begin{figure}[h]
\begin{center}
\includegraphics[width=5cm]{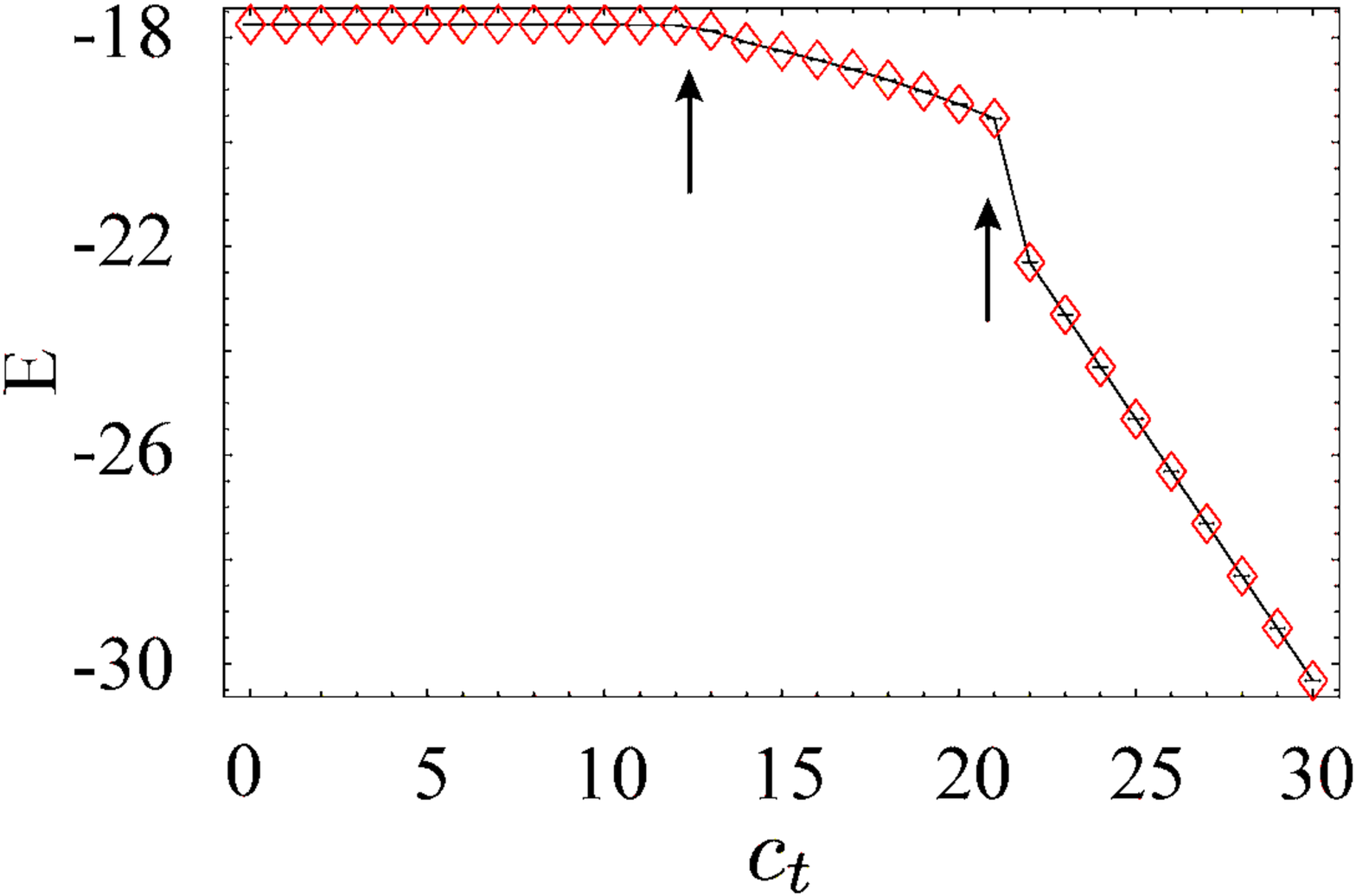}
\hspace{0.5cm}
\includegraphics[width=5cm]{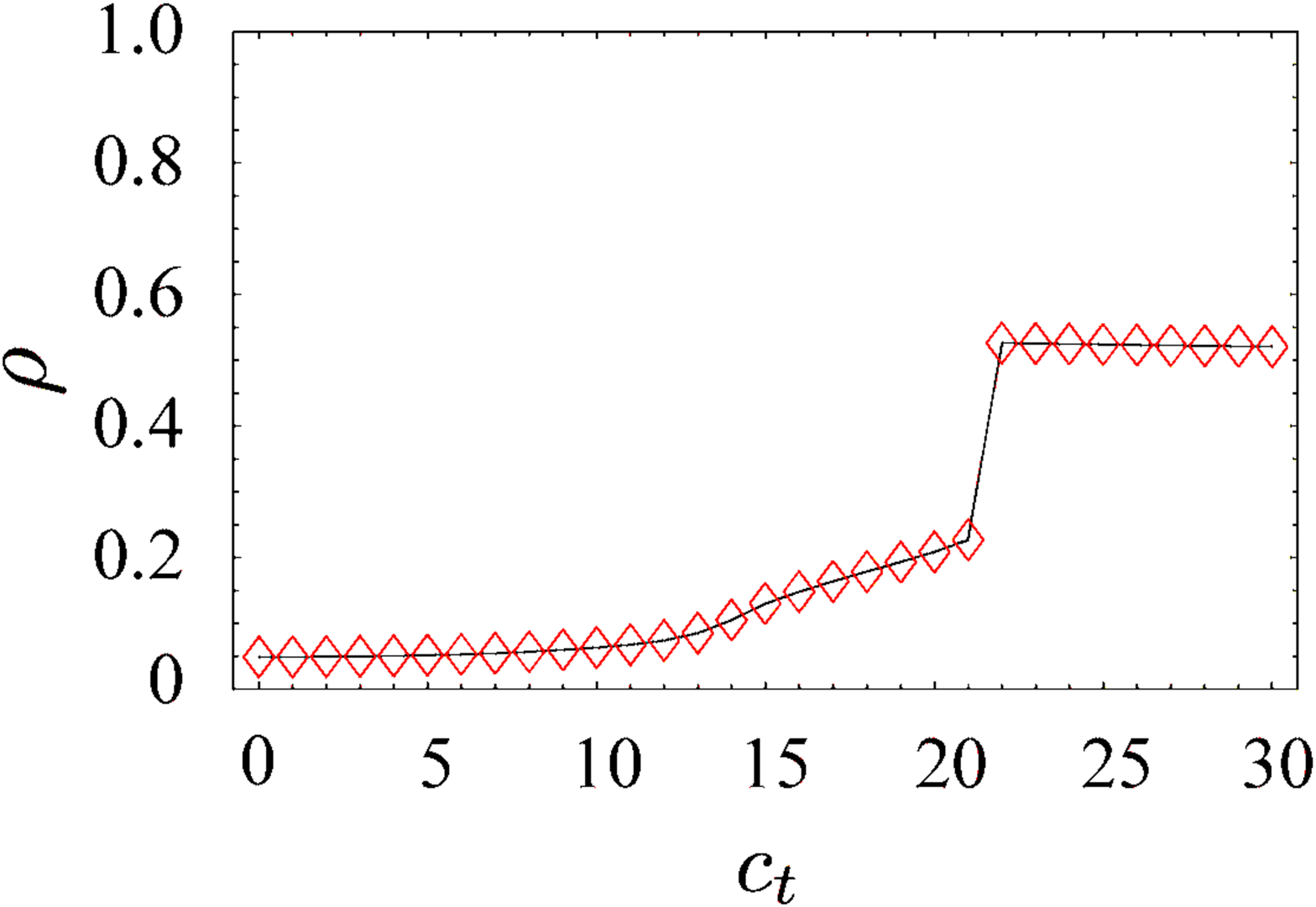}
\vspace{-0.3cm}
\caption{
Behavior of internal energy and hole density at phase boundaries of supersolid
indicating phase transitions at $c_t \simeq 12.0$ and $c_t \simeq 20.0$,
where $\beta\mu=5$.
Arrows indicate the location of phase transitions.
}\vspace{-0.5cm}
\label{SS}
\end{center}
\end{figure}

\begin{figure}[h]
\begin{center}
\includegraphics[width=5cm]{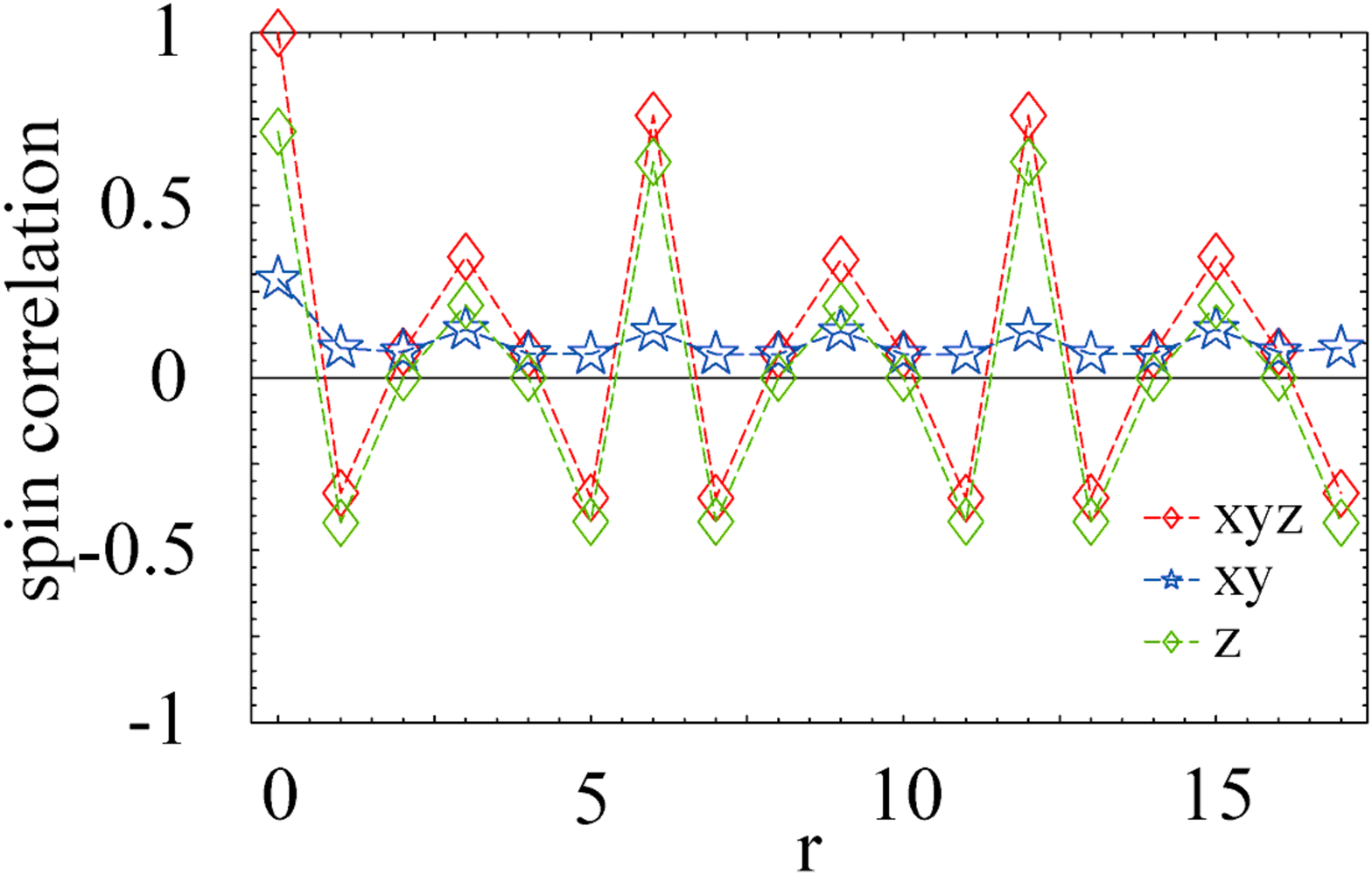}
\hspace{0.5cm}
\includegraphics[width=5cm]{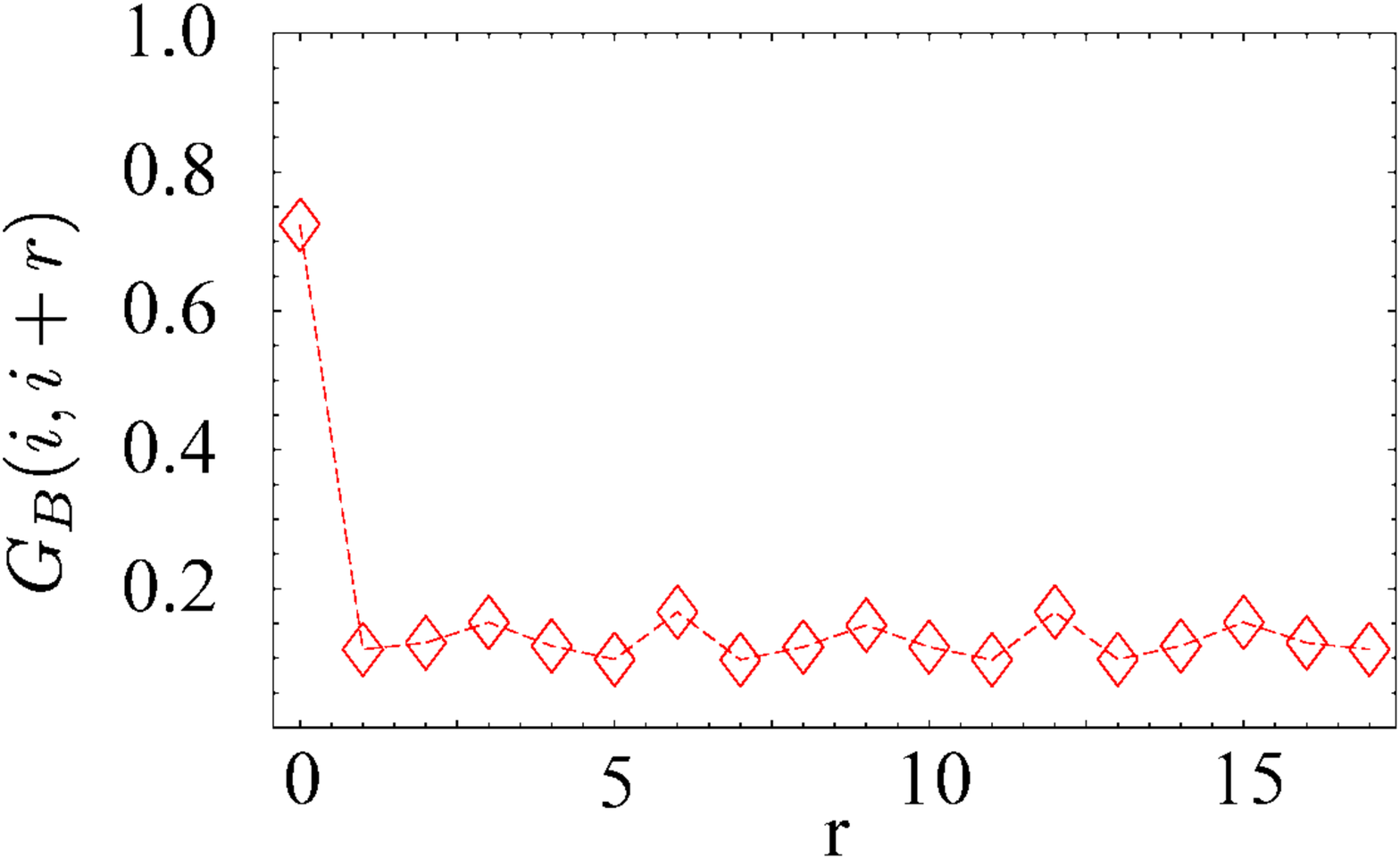}
\vspace{-0.3cm}
\caption{
Spin and particle correlation functions in supersolid.
$c_t=17$ and $\beta\mu=5$.
}\vspace{-0.5cm}
\label{Corr_SS}
\end{center}
\end{figure}

The obtained phase diagram by the MC simulations is shown in Fig.\ref{PD4}.
Phase transition from the pure-spin state to the supersolid is of
second order and the others are of first order.
The locations of the phase boundary
are determined by the measurement of $E$ and $C$.
As we expected, there is a parameter region in which the supersolid forms.
The behavior of the internal energy and hole density are shown in
Fig.\ref{SS} at the phase boundary of the supersolid.
Correlation functions to be used for identify the supersolid are explicitly shown
in Fig.\ref{Corr_SS}.
The correlation of the $z$-component of spin indicates there exists
solid order, and the particle correlation shows the existence of a small but finite
long-range order, i.e., superfluidity.

\section{Conclusion and discussion}

In this paper we studied phase diagram of the bosonic t-J model 
in the stacked triangular lattice.
The model has a rich phase structure and we expect that some of them
are observed by experiment on systems of two-component cold atomic gas
and strongly correlated electron systems.

\begin{figure}[h]
\begin{center}
\includegraphics[width=7cm]{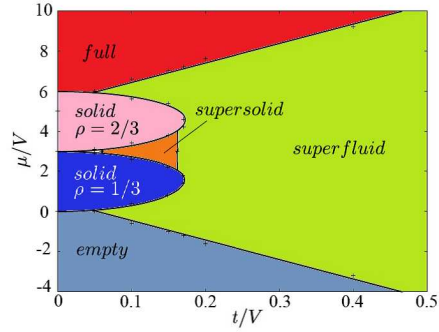}
\vspace{-0.3cm}
\caption{
Phase diagram of the hard-core boson system in the stacked triangular
lattice.\cite{SS3D}
In this figure, $\rho$ {\em denotes the particle density}.
$V$ and $t$ are parameters of the repulsion and hopping, and we set
$\beta V=50$ (low-$T$ region).
There exists a supersolid state between two solid states.
}\vspace{-0.5cm}
\label{HCB}
\end{center}
\end{figure}
\begin{figure}[h]
\begin{center}
\includegraphics[width=7cm]{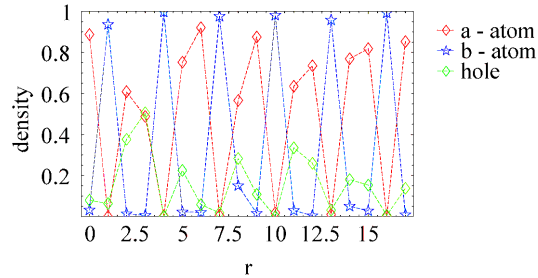}
\vspace{-0.3cm}
\caption{
Density profiles of $a$, $b$-atoms and hole in the supersolid.
$c_{Jz}=10.0,\ c_t=17.0$ and $\beta\mu=5.0$.
}\vspace{-0.5cm}
\label{density_tJz}
\end{center}
\end{figure}
It is interesting to compare the obtained phase diagrams with that of the
hard-core boson system in Ref.\cite{SS3D}, which is shown in Fig.\ref{HCB}.
In the model Hamiltonian,
$V$ represents the nearest-neighbor repulsive interaction and
$t$ is the hopping parameter.
There are two solid states with the particle density $\sim 1/3$ and $2/3$, 
and a supersolid state between them.
The phase diagram in Fig.\ref{HCB} respects the particle-hole symmetry
of the model, whereas in the t-J model such a symmetry does not exist.
In the t-J$_{xy}$ model studied in the paper, the supersolid does not
form in similar parameter region and the phase separation takes place there instead,
whereas in the  t-J$_z$ model the supersolid appears as in the hard-core
boson system.
For  t-J$_{xy}$ model with $t>0$,
the reason why the phase-separated state dominates was explained in Sec.3.2.
The state shown in Fig.\ref{1/3_grand} corresponds to the state with the
particle density $=2/3$ in the hard-core boson system.
In the t-J model, the state with particle density $=1/3$ generally does not
have spin long-range order nor solid (density-wave) order because of the
low particle density and the non-existence of the particle-hole symmetry.
Structure of the supersolid in the t-J$_z$ model is slightly different from
that observed in the hard-core boson system.
The particle density there is larger than $2/3$, and one kind of atom, e.g., 
$b$-atoms occupy $B$-sublattice and $a$-atoms and holes form a superposed
state in $A$ and $C$-sublattices.
See Fig.\ref{density_tJz}.
Therefore for appearance of the supersolid in the t-J$_z$ model, 
spin degrees of freedom plays an essential role.
On the other hand in the hard-core bosons on the triangular lattice,
the supersolid can be interpreted as a superfluid of excess bosons (or holes)
via a relay-type movement.

It is interesting and useful to discuss the relation between the bosonic t-J model
studied in this paper and the fermionic t-J model on the triangular lattice
whose phase diagram was studied by high-temperature expansion in Ref.\cite{ogata}.
For the Hubbard model on the square lattice, very recent study by the
numerical link-cluster expansion shows that the Fermi-Hubbard and 
Bose-Hubbard have a very similar physical properties for large on-site
repulsions\cite{FBHubbard}.
On the other hand in Ref.\cite{ogata}, the authors studied the fermionic
t-J model on the triangular lattice and concluded that 
the RVB state appears by hole doping into the state with the three-sublattice
spin symmetry for $t>0$ and $\rho=0.2-0.6$.
By the high-temperature expansion, they found that the entropy decreases
considerably at low temperature and peak of the spin susceptibility
moves to the high-temperature region.
These results indicate that the hole doping releases the frustration and
enhances AF correlation though AF long-range order is not observed.
In their discussion on the realized state, it was assumed that 
the low-temperature state is homogeneous and metallic.
If this assumption is correct, the conclusion that the RVB forms
at low-temperature seems plausible and acceptable.
However the results obtained in this paper clearly offers another
interpretation of their calculations, i.e., the state shown in Fig.\ref{1/3_grand}
has obviously has lower entropy and stable AF correlation compared to
the 120$^o$ spin state at $\rho=0$.
Holes are localized there and the state is inhomogeneous contrary to the
assumption in Ref\cite{ogata}.
We expect that a similar state to that in Fig.\ref{1/3_grand} is
realized in the fermionic model, though the superfluid in Fig.\ref{PD2}
corresponds to a metallic phase of mobile holes in the fermionic
t-J model.
On the other hand for $t<0$, it was found in Ref.\cite{ogata}
that the entropy remains large until very low temperature and therefore
the hole doping does not release the spin frustration at $\rho=0$.
This result is also consistent with the results shown, e.g., in
Fig.\ref{correlation3}.

In the above discussion on the bosonic and fermionic t-J models on the
triangular lattice, we directly compared the obtained results for the
models, and concluded that there is a close resemblance between
two models.
It is important and interesting to show the relationship between two
model by an analytical method.
To this end, discussion using the Chern-Simons gauge theory in Ref.\cite{CS}
is useful.
In the slave-particle representation (\ref{slave1}), the Hamiltonian (\ref{HtJ})
is expressed as 
\begin{eqnarray}
H_{\rm tJ} &=& -t \sum_{\langle i, j\rangle, \sigma}
(\varphi^\dagger_{\sigma i}\phi_i\phi^\dagger_j \varphi_{\sigma j}+\mbox{h.c.})
 \nonumber \\
&& +J\sum_{\langle i, j\rangle, \sigma, \sigma'}
(\epsilon_{\sigma\bar{\sigma}}
\varphi^\dagger_{\bar{\sigma}i}\varphi^\dagger_{\sigma j}
\epsilon_{\sigma'\bar{\sigma}'}
\varphi_{\sigma' j}\varphi_{\bar{\sigma}' i}+\mbox{h.c.})P_iP_j,
\label{HtJ2}  \\
&& P_i=1-\phi^\dagger_i\phi_i,
\end{eqnarray}
where $\bar{\sigma}=2(1)$ for $\sigma=1(2)$, 
$\epsilon_{12}=-\epsilon_{21}=1, \ \epsilon_{11}=\epsilon_{22}=0$, 
and we have set $J_z=J$
for simplicity.
On the other hand, the Hamiltonian of the fermionic t-J model in two
spatial dimensions is given as
follows by using the Chern-Simons gauge fields $A^h_{ij}$ and
$A^s_{ij}$ defined on the link $(i,j)$\cite{CS,IMCS},
\begin{eqnarray}
H^{\rm F}_{\rm tJ} &=& -t \sum_{\langle i, j\rangle, \sigma}
(e^{iA^s_{ij}}e^{-i\sigma A^h_{ij}}
\varphi^\dagger_{\sigma i}\phi_i\phi^\dagger_j \varphi_{\sigma j}+\mbox{h.c.})
\nonumber \\
&& +J\sum_{\langle i, j\rangle, \sigma, \sigma'}
(e^{-i\sigma A^h_{ij}}\epsilon_{\sigma\bar{\sigma}}
\varphi^\dagger_{\bar{\sigma}i}\varphi^\dagger_{\sigma j}
e^{i\sigma' A^h_{ij}}\epsilon_{\sigma'\bar{\sigma}'}
\varphi_{\sigma' j}\varphi_{\bar{\sigma'} i}+\mbox{h.c.})P_iP_j,
\label{HftJ}
\end{eqnarray}
where $\sigma A\equiv A \ (-A)$ for $\sigma=1 \ (2)$, and
\begin{eqnarray}
&& \sum_{(i,j) \in {\cal C}} A^s_{ij}=\pi \sum_{(i,j) \ {\rm inside} \ {\cal C}}
(\varphi^\dagger_1\varphi_1-\varphi^\dagger_2\varphi_2), \; \mbox{mod} \ 2\pi,
\nonumber \\
&& \sum_{(i,j) \in {\cal C}} A^h_{ij}=\pi \sum_{(i,j) \ {\rm inside} \ {\cal C}}
(\phi^\dagger \phi-\rho), \; \mbox{mod} \  2\pi,
\label{CSgauge}
\end{eqnarray}
where ${\cal C}$ is a closed loop and $A_{ij}=-A_{ji}$ is understood 
in the above summations in (\ref{CSgauge}).
In the homogeneous state, we can replace the RHS's of Eq.(\ref{CSgauge})
by their mean values for investigating the ground-state phase diagram.
In the case of the coplanar spin configuration as we considered in this
paper, effects of $A^s_{ij}$ is small because 
$\langle (\varphi^\dagger_1\varphi_1-\varphi^\dagger_2\varphi_2) \rangle=0$.
Similarly for a homogeneous hole distribution,
the Chern-Simons gauge field $A^h_{ij}$ can be set $A^h_{ij}=0$ in the mean-field
level.
Therefore for a homogeneous state, phase diagrams of $H_{\rm tJ}$ and 
$H^{\rm F}_{\rm tJ}$ are closely related with each other, 
though dispersion relation of excitations in the two systems are different by 
the local constraints Eq.(\ref{CSgauge}),
i.e., in the system $H^{\rm F}_{\rm tJ}$, hopping of bosonic spinon and 
holon accompanies the Chern-Simons flux\cite{CS2}.
Anyway, more detailed study on the relation between the two models by using the 
Chern-Simons gauge theory is interesting and useful\cite{CS3}.


\begin{acknowledgment}
This work was partially supported by Grant-in-Aid
for Scientific Research from Japan Society for the 
Promotion of Science under Grant No23540301.
\end{acknowledgment}

\end{document}